\documentclass[10pt]{iopart}

\usepackage{iopams}
\usepackage{amssymb}
\usepackage{amsfonts}
\usepackage{amstext}
\usepackage{graphicx}
\usepackage{setstack}
\usepackage{amscd}
\usepackage{mathrsfs}
\usepackage{stmaryrd}

\renewcommand{\Im}{\mathrm{Im}}
\newcommand{\Te}{\mathbb{T}e}
\newcommand{\Pe}{\mathbb{P}e}

\newcommand{\rrangle}{\rangle \hspace{-0.2em} \rangle}

\newcommand{\rrrangle}{\rangle \hspace{-0.2em} \rangle \hspace{-0.2em} \rangle}

\newcommand{\id}{\mathrm{id}}
\newcommand{\xrightarrow}[1]{\overset{#1}{\longrightarrow}}
\newcommand{\yrightarrow}[1]{\underset{#1}{\longrightarrow}}
\newcommand{\A}{\mathbf A}
\newcommand{\B}{\mathbf B}
\newcommand{\lambdabar}{\mkern0.75mu\mathchar '26\mkern -9.75mu\lambda}

\begin{document}

\title{Adiabatic transport of qubits around a black hole}

\author{David Viennot \& Olivia Moro}
\address{Institut UTINAM (CNRS UMR 6213, Universit\'e de Bourgogne-Franche-Comt\'e, Observatoire de Besan\c con), 41bis Avenue de l'Observatoire, BP1615, 25010 Besan\c con cedex, France.}

\begin{abstract}
We consider localized qubits evolving around a black hole following a quantum adiabatic dynamics. We develop a geometric structure (based on fibre bundles) permitting to describe the quantum states of a qubit and the spacetime geometry in a single framework. The quantum decoherence induced by the black hole on the qubit is analysed in this framework (the role of the dynamical and geometric phases in this decoherence is treated), especially for the quantum teleportation protocol when one qubit falls to the event horizon. A simple formula to compute the fidelity of the teleportation is derived. The case of a Schwarzschild black hole is analysed. 
\end{abstract}


\section{Introduction}
Recent works have explored the possibility to show the effects of the gravity onto quantum systems \cite{Ahmadi,Brushi}. An interesting study concerning the behaviour of a scalar field in the neighbourhood of a black hole \cite{Fuentes} has shown that the entanglement is degraded by the effect of the black hole. In this work, I. Fuentes-Schuller and R.B. Mann have studied a model in which all the ingredients of quantum field theory are present, but the gravitation is only represented as a Rindler spacetime corresponding to the uniform surface gravity in the neighbourhood of the event horizon. Recently M.C. Palmer {\it etal} \cite{Palmer} have proposed a theory of localized qubits in curved spacetimes. In this paper, we want to reexamine in this framework the problem of qubits around a black hole, especially concerning the entanglement and the quantum teleportation protocol. Moreover we want to analyse the physical meaning of the non-self-adjointness of the localized qubit Hamiltonian. In contrast with the model of Fuentes-Schuller and Mann, quantum field theory is not completely treated, semi-classical approximations in the localized qubit theory induce the lost of the possibility to create and annihilate particles, the lost of the Unruh effect, and the lost of the delocalization of the wave packets (but the non-locality remains in the theory with the entanglement). But we want to treat the complete geometry of the black hole spacetime. The interest of our approach  is the possibility to treat the local effects on a fixed quantum system (a single qubit boarded in a ``spacecraft'' with definite position and velocity and following a geodesic) whereas no local information associated with the spacetime geometry is taken into account in the Fuentes-Schuller Mann model since the gravitational field is considered in it as uniform and the qubits are completely delocalized in it. We want to analyse the entanglement degrading effect with respect to the position and the velocity of the qubits with respect to the black hole.\\
We consider a qubit realised as the spin of a fermion submitting only to the gravitational field (no external magnetic field) and boarded in a ``spacecraft'' following a geodesic around a black hole. By ``gravitational field'' we mean the Lorentz connection associated with the spacetime geometry. In order to treat the dynamics of the qubit, we use the quantum adiabatic approximation because the qubit transport can be considered as slow with respect to its proper quantum time (the period of its Rabi oscillations induced by the gravitational field). Section 2 summarizes the localized qubit approach by rewritting it in the language of the fibre bundle theory. The goal of this reformulation is to provide a description including spacetime geometries and qubit quantum states in a single common geometric structure. This is achieved by the introduction of the fiber bundles of the quantum adiabatic approximation. We show that the problem takes place in complex line bundles over a space constituting by the product of the space of Lorentz connections by the tangent bundle of the spacetime manifold. We show that the qubit is submitted to a kind of decoherence process induced by the gravitational field and responsible to the degradation of the entanglement. Section 3 presents in our framework the quantum teleportation protocol with an EPR (Einstein Podolsky Rosen) qubit pair when one qubit falls to the black hole whereas the other one is comoving with it. We compute a formula providing the fidelity of the teleportation. Section 4 applies the formalism to two spacetime geometries, firstly to the Rindler spacetime used by Fuentes-Schuller and Mann, secondly to a Schwarzschild spacetime, where we analyse the fidelity of the quantum teleportation protocol with respect to the geodesic followed by the qubit falling to the black hole.\\

{\it Throughout this paper, we consider the unit system such that $\hbar=c=1$.}\\

{\it Note about the notations: a fibre bundle of total space $FB$ and base space $B$ is denoted by its projection $FB \to B$. The space of the local sections of a fibre bundle is denoted by $\Gamma(B,FB)$. For a map $f:B' \to B$, $f^*:FB \to FB'$ denotes the map induced by the fibration. Let $M$ be a manifold, its tangent bundle is denoted by $TM$ ($T_xM$ is the tangent space of $M$ at the point $x$), and its differential $1$-form set is denoted by $\Omega^1 M$. For a map $f:M \to N$, $f_*:TM \to TN$ and $f^*:\Omega^1 N \to \Omega^1 M$ denote the associated tangent and the cotangent maps (the pusch-forward and the pull-back). $\mathscr C^0(M)$ and $\mathscr C^\infty(M)$ denote the spaces of continuous and differentiable functions of $M$. ''$\simeq$'' between two manifolds denotes a diffeomorphism. $\Pr_i : E_1 \times ... \times E_n$ denotes the projection map defined by $\Pr_i(e_1,...,e_n) = e_i$. We use the Einstein convention concerning the up and down indices repetition. The greek indices runs in $\{0,1,2,3\}$ as curved spacetime indices, the capital latin indices runs in $\{0,1,2,3\}$ as flat Minkowski auxiliary spacetime indices, the small latin indices runs in $\{1,2,3\}$. $\dot x^\mu = \frac{dx^\mu}{d\tau}$ denotes the derivation with respect to a proper time $\tau$.}

\section{Adiabatic dynamics of a localized qubit}
\subsection{Localized qubit in a curved spacetime}
In this section we summarize the results (without details) of Palmer \textit{etal} \cite{Palmer}, we fix the notations, and we embbed the localized qubit theory into the fibre bundle theory (for an exposition of the fibre bundle theory see for example \cite{Nakahara}). Let $M$ be an open set of the spacetime manifold endowed with a local coordinates system $\{x^\mu\}_{\mu\in\{0,1,2,3\}}$ and a metric tensor $g_{\mu \nu}(x)$ (to simply we will refere to $M$ as the spacetime). Let $\{e^A_\mu(x)\}_{\mu,A\in\{0,1,2,3\}}$ be a tetrad field associated with the metric: $g_{\mu \nu} = \eta_{AB} e^A_\mu e^B_\nu$, where $\eta_{AB}$ is the Minkowski metric. Let $\omega_\rho^{AB} = e^A_\mu \partial_\rho e^{B\mu} + e^A_\mu \Gamma^\mu_{\rho \nu} e^{B\nu}$ be the Lorentz connection ($\Gamma^\mu_{\rho \nu}$ being the Christoffel symbols). A Dirac field $\Psi$ obeys to the Dirac-Einstein equation:
\begin{equation}
  \label{DiracEinstein}
(\imath \gamma^A e_A^\mu(x) \nabla_\mu - m) \Psi(x) = 0
\end{equation}
where $\{\gamma^A\}_{A\in\{0,1,2,3\}}$ are the Dirac matrices (in Weyl representation) and $\nabla_\mu$ is the spinorial covariant derivative defined by
\begin{equation}
\nabla_\mu = \frac{\partial}{\partial x^\mu} + \omega_\mu^{AB}(x) \mathfrak D(M_{AB})
\end{equation}
where $\mathfrak D$ is the $(1/2,0)\oplus(0,1/2)$ representation of $SL(2,\mathbb C)$ (the coverging group of the Lorentz group $SO^+(3,1)$) on $\mathbb C^2 \oplus \mathbb C^2$ (we denote by the same symbol the induced representation of its Lie algebra), i.e. $\mathfrak D(M_{AB}) = \frac{1}{4}[\gamma_A,\gamma_B]$.\\
Let $TM \to M$ be the tangent bundle of $M$ and $FM \to M$ be the frame principal $SO^+(3,1)$-bundle of $M$. Let $\varphi_T : M \times \mathbb R^4 \xrightarrow{\simeq} TM$ and $\varphi_F : M \times SO^+(3,1) \xrightarrow{\simeq} FM$ be the local trivializations of these bundles, which are defined by $\varphi^\mu_{T} (x,v) = e^\mu_A(x) v^A$ and $\varphi_F(x,\Lambda) = e(x)\Lambda$ ($e \in GL(4,\mathbb R)$ is the matrix of elements $e^\mu_A$). The tetrad field can be identified as the trivializing local section of $FM$: $x\mapsto e(x)=\varphi_F(x,\id) \in \Gamma(M,FM)$. Let $P \to M$ be the principal $SL(2,\mathbb C)$-bundle associated with the local $SL(2,\mathbb C)$ transformations of the spinors ($P$ is an extension of $FM$ such that $FM=P/\mathbb Z_2$). Let $E \to TM$ and $\bar E \to TM$ be the associated vector bundles for the representation $(1/2,0)$ and $(0,1/2)$, i.e. $E \to TM$ is defined by its local trivialization $\varphi_E : TM \times \mathbb C^2 \xrightarrow{\simeq} E$, $\varphi_E(v(x),\psi) = [\varphi_P(\pi_T(v(x)),g),g^{-1} \psi]_{g \in SL(2,\mathbb C)}$ where $\varphi_P$ is the local trivialization of $P \to TM$ and $\pi_T$ is the projection $TM \to M$, and $\varphi_{\bar E}(v(x),\psi) = [\varphi_{\bar P}(\pi_T(v(x)),g),\bar g^{-1} \psi]_{g \in SL(2,\mathbb C)}$ (we have denoted simply the $(1/2,0)$-action of $g \in SL(2,\mathbb C)$ on $\psi \in \mathbb C^2$ by $g\psi$ and the $(0,1/2)$-action of $g$ on $\psi$ by $\bar g \psi$).\\
$\Gamma(TM,E\oplus \bar E)$ is a Hilbert $\mathscr C^0(TM)$-module endowed with the inner product:
\begin{eqnarray}
& & \forall \Psi,\Phi \in \Gamma(TM,E\oplus \bar E),\nonumber \\
& & \quad \langle \Psi|\Phi\rangle_{\Gamma(TM,E\oplus \bar E)}(u(x)) = \langle \Psi(u(x))|\gamma^0\gamma^A |\Phi(u(x))\rangle_{\mathbb C^4} u_A(x)
\end{eqnarray}
where $u(x) \in T_xM$ ($u^A u_A = 1$, $u_A = e_A^\mu u_\mu$). Let $\Sigma \subset M$ be a spacelike hypersurface of $M$ and $N^+ \Sigma = \{n \in TM_{|\Sigma} ; \forall t \in T\Sigma, g_{\mu \nu} n^\mu t^\nu =0, g_{\mu \nu} n^\mu n^\nu > 0\}$ be the set of futur oriented timelike normal vectors to $\Sigma$. The Dirac field $\Psi$ is a vector of the Hilbert space $L^2(N^+\Sigma,E\oplus \bar E) = \{\Psi \in \Gamma(N^+\Sigma,E\oplus \bar E); \int_\Sigma \|\Psi\|^2_{\Gamma(TM,E\oplus \bar E)} (n(x)) d\Sigma(x) < +\infty\}$ (the space of the Dirac spinor fields living at some time on the instantaneous space $\Sigma$).\\

Let $\mathcal C$ be a geodesic worldline in $M$ and $\{\Sigma_\tau\}_{\tau \in \mathbb R}$ be a fiolation of $M$ along $\mathcal C$ by spacelike hypersurfaces ($\tau$ being the proper time along $\mathcal C$). By the WKB (Wentzel Kramers Brillouin) approximation associated with the assumption that the Compton wavelength is small with respect to the curvature scale \cite{Palmer}, we have $\int_{\mathbb R}^\oplus L^2(N^+\Sigma_\tau,E\oplus \bar E)d\tau \xrightarrow{WKB} \Gamma(T\mathcal C,E \oplus \bar E)$. This semi-classical approximation suppresses the space delocalisation of the fermion supporting the qubit and the absence of second quantization supresses the particle number ambiguity. Since only the $(1/2,0)$-representation is needed to describe a single spin (qubit), we can project onto the space $\Gamma(T\mathcal C,E)$. We work then with the composite bundle \cite{Viennot} $E \to TM \to M$. It can be more easy to work with a bundle $E_+ \to M$ with structure fibre $\mathbb R^4 \times \mathbb C^2$ (tangent vector model space and spin quantum state space). $E_+$ is defined by the local trivialization $\varphi_{E+} : M \times \mathbb R^4 \times \mathbb C^2 \xrightarrow{\simeq} E_+$, $\varphi_{E+}(x,v,\psi) = \varphi_E(\varphi_T(x,v),\psi) = \varphi_E(e(x)v,\psi)$. We have an action of $SL(2,\mathbb C)$ onto $E_+$ defined by $\forall g \in SL(2,\mathbb C)$, $\mathfrak D_+(g) \varphi_{E+}(x,v,\psi) = \varphi_{E+}(x,\Lambda(g)v,\mathfrak D(g)\psi)$, where $\Lambda : SL(2,\mathbb C) \to SO^+(3,1)$ is the group homomorphism associated with the quotient $SO^+(3,1) \simeq SL(2,\mathbb C)/\mathbb Z_2$.\\
We can identify the space of local sections of $E_+$, $\Gamma(M,E_+)$, to the space of $SO^+(3,1)$-invariant local sections of $E$: $\Gamma_i(TM,E) =\{\psi \in \Gamma(TM,E); \forall \Lambda \in SO^+(3,1), \forall v \in TM, \psi(\Lambda v(x)) = \psi(v(x))\}$. The restriction of $\Gamma(TM,E)$ to the invariant sections is important to ensure the physical character of the theory, more precisely to have the following property:
\begin{eqnarray}
& & \forall \psi,\phi \in \Gamma_i(TM,E), \forall g \in SL(2,\mathbb C), \forall v \in TM \nonumber \\
& & \quad \langle \mathfrak D(g) \psi|\mathfrak D(g) \phi\rangle_{\Gamma(TM,E)}(\Lambda(g)v(x)) = \langle \psi|\phi\rangle_{\Gamma(TM,E)}(v(x))
\end{eqnarray} 
i.e. the quantum propabilities are invariant under Lorentz transformations. Endowed with the inner product $\langle \psi_v | \phi_w \rangle_{\Gamma(M,E^+)}(x) = \langle \psi|\phi\rangle_{\Gamma(TM,E)} (v(x)) \delta(v(x)-w(x))$ (with $\psi_v(x)=\psi(v(x)) = \varphi_E(\Pr_{2,3}\varphi_{E+}^{-1}(\psi_v))$, $\delta$ is the Dirac distribution), $\Gamma(M,E_+)$ is a Hilbert $\mathscr C^0(M)$-module.\\

Concerning the evolution of the Dirac field, by using the Weyl representation of the Dirac matrices, the Dirac-Einstein equation \ref{DiracEinstein} can be rewritten as the Van der Waerden equation:
\begin{equation}
 e^\mu_A \sigma^A e^\nu_B \bar \sigma^B \nabla_\mu \nabla_\nu \Psi_E + m^2 \Psi_E = 0
\end{equation}
with $\Psi_E \in \int_{\mathbb R}^\oplus L^2(N^+\Sigma_\tau,E)d\tau$ and with $\{\sigma^A\}_A = \{\id,\sigma^x, \sigma^y, \sigma^z\}$ and $\{\bar \sigma^A\}_A = \{\id,-\sigma^x, -\sigma^y, -\sigma^z\}$ ($(\sigma^x,\sigma^y,\sigma^z)$ being the usual Pauli matrices). After some algebra (see \cite{Palmer}) it can be rewritten as
\begin{equation}
\label{VanderWaerden}
g^{\mu \nu} \nabla_\mu \nabla_\nu \Psi_E - \imath e_\mu^A e_\nu^B L_{A B} \mathfrak R^{\mu \nu} \Psi_E + m^2 \Psi_E = 0
\end{equation}
with $L_{AB} = \frac{\imath}{4}(\sigma^A \bar \sigma^B - \sigma^B \bar \sigma^A)$ and $\mathfrak R^{\mu \nu}$ the Ricci tensor. The WKB ansatz consists to set $\Psi_E = \psi e^{\imath S}$ with $\nabla_\mu S = k_\mu$ ($k_\mu$ is the wave number, $k^\mu k_\mu = m^2$) and the WKB approximation consists to assume that the typical scale over which $\psi$ varies and the spacetime curvature scale are large compared to the Compton wavelength $\lambdabar = \frac{1}{m}$. Inserting the WKB representation of $\Psi_E$ into equation \ref{VanderWaerden} and neglecting the small terms with respect to the WKB approximation, we find (see \cite{Palmer})
\begin{equation}
  2k^\mu \nabla_\mu \psi + \psi \nabla_\mu k^\mu = 0
\end{equation}
Finally, along the geodesic $\mathcal C$ defined by $\ddot x^\mu + {\Gamma^\mu}_{\nu \rho} x^\nu x^\rho =0$, with $\frac{k^\mu}{m} = \dot x^\mu$, the localized qubit is described by the spin state $\psi \in \Gamma_i(TM,E) \simeq \Gamma(M,E_+)$ which obeys to the Schr\"odinger like equation (see \cite{Palmer}):
\begin{equation}
\label{SchrodingerLocalized}
\imath \frac{d\psi}{d\tau} = - \frac{1}{2} \omega^{AB}_\mu(x(\tau)) \dot x^\mu(\tau) L_{AB} \psi(\tau)
\end{equation}
where $\tau$ is the proper time along the geodesic followed by the qubit. The WKB approximation is equivalent to say that $\psi$ consists to a wave packet essentially localized around the geodesic, with a wave packet size very small with respect to the spacetime curvature scale. According to this remark, we can consider that the qubit viewed at the curvature scale, is localized on the geodesic (a more complete discussion about the qubit localization and the WKB approximation can be found in \cite{Palmer}).\\
$\psi(\tau) \in \pi_E^{-1}(u(\tau))$ where $\pi_E$ is the projection of $E$ onto $TM$ and $u(\tau) = \dot x^\mu(\tau) \frac{\partial}{\partial x^\mu} \in T_{x(\tau)}M$. The instantaneous qubit Hilbert space  $\pi_E^{-1}(u(\tau)) \subset \Gamma_i(TM,E)$ depends on the four-velocity since the correct inner product induced by the Dirac field theory is
\begin{eqnarray}
\langle \psi | \phi \rangle_{\Gamma(TM,E)}(u(x)) & = & \langle \psi(u(x))|\bar \sigma^A | \phi(u(x)) \rangle_{\mathbb C^2} u_A(x) \\
& = & \langle \psi^\star(u(x))|\phi(u(x))\rangle_{\mathbb C^2}
\end{eqnarray}
where the conjugate state is $\psi^\star(u(x))= \bar \sigma^{A\dagger} u_A(x) \psi(u(x))$.\\

The interaction of the Dirac field with gravity is essentially encoded in the Dirac-Einstein equation \ref{DiracEinstein} by the spinorial covariant derivative. We refind this term in the localized qubit Schr\"odinger equation \ref{SchrodingerLocalized} with the operator $-\frac{1}{2}  \omega_\mu^{AB} \dot x^\mu L_{AB}$. We see that equations \ref{DiracEinstein} and \ref{SchrodingerLocalized} are similar to equations of a fermion or a spin with minimal coupling to an interaction field having the Lorentz connection as gauge potential. We can then consider the Lorentz connection as the ``gravity field'' felt by the fermion or the qubit. More subtly, another aspect of general relativity is encoded in the Dirac-Einstein equation \ref{DiracEinstein} by the tetrad field. $\{e^A(x)\}_{A}$ defines the local inertial frame in the neighbourhood of $x$. The Dirac-Einstein equation is based on the idea that the equation of the field takes in this frame a similar form of its equivalent into a Minkowski spacetime. In the localized qubit theory, we refind this dependency with the frame by the dependency of the Hilbert space (and of its inner product) with the four-velocity. The physical meaning of this dependency will be extensively discussed in the sequel of this paper.

\subsection{Inner products and linear functionals}
In order to interpret the dependence of the instananeous spin Hilbert space on the four-velocity, it needs to recall some elementary axioms of the quantum mechanics. An Hilbert space $\mathcal H$ constitutes the space of the states of a quantum system. But its algebraic dual $\mathcal H^*$, i.e. the space of continuous linear functionals of $\mathcal H$, is the space of the probability amplitudes of the elementary events: $\ell \in \mathcal H^*$ is a map from $\mathcal H$ to $\mathbb C$, such that $|\ell(\psi)|^2$ is the probability of the realization of some measurement event associated with $\ell$ when the quantum system is in the state $\psi$. By the Riesz theorem, we know that $\forall \ell \in \mathcal H^*$, $\exists !\eta_\ell \in \mathcal H$ (up to a renormalization and phase factor) such that $\forall \psi \in \mathcal H$, $\ell(\psi) = \frac{\langle \eta_\ell|\psi \rangle_{\mathcal H}}{\|\eta_\ell\|_{\mathcal H} \|\psi\|_{\mathcal H}}$. We find $\langle \eta_\ell| \in \mathcal H^*$ as an eigenvector of an observable $\Theta$ associated with the measure; $\langle \eta_\ell|$ corresponds to the event ``the measure of $\Theta$ has provided the result $\lambda_\ell$'' ($\lambda_\ell$ being the eigenvalue associated with the eigenvector $\langle \eta_\ell|$). It is then important to note that a ``ket'' $|\psi\rangle$ characterizes the quantum system as being its state, whereas a ``bra'' $\langle \eta|$ characterizes an event for an observer making measures on the quantum system.\\
Returning to the localized qubit problem, we want to interpret the difference between the two linear functionals $\langle \psi|$ and $\langle \psi^\star| = \langle \bar \sigma^\dagger_A u^A \psi|$ (the ``bra'' always denoting in this paper the partial inner product of $\mathbb C^2$: $\langle \psi|.\rangle_{\mathbb C^2}$, and never $\langle \psi|.\rangle_{\Gamma(TM,E)}$). Let $\psi \in \Gamma_i(TM,E)$ be a normalized state:
\begin{eqnarray}
\langle \psi|\psi \rangle_{\Gamma(TM,E)} = 1 & \iff & \langle \psi|\bar \sigma^A|\psi\rangle_{\mathbb C^2} u_A = 1 \\
& \iff & \|\psi\|^2_{\mathbb C^2} u_0 - \langle \psi|\sigma^i|\psi \rangle u_i = 1 \\
\label{spinchangeref}
& \iff & \gamma \left(S^0 - \vec S \cdot \vec v \right) = S^{0\star}
\end{eqnarray}
where we have introduced the magnetic four-momentum operator $\{\hat S^A\}_A = \{\frac{1}{2} \id, \frac{1}{2} \sigma^x,\frac{1}{2} \sigma^y,\frac{1}{2} \sigma^z\}$, with $S^A = \langle \psi|\hat S^A|\psi \rangle_{\mathbb C^2}$; $\gamma=u_0$, $\gamma \vec v = \vec u$ and $S^{0\star} = \langle \psi|\frac{\id}{2}|\psi \rangle_{\Gamma(TM,E)} = \langle \psi^\star|\frac{\id}{2}|\psi\rangle_{\mathbb C^2} = \frac{1}{2}$. The formula \ref{spinchangeref} is the classical relation between a magnetic four-momentum $S$ measured into an inertial frame $K$ and the four-momentum $S^\star$ measured into its rest frame $K^\star$ of four-velocity $\vec u = (\gamma,\gamma \vec v)$ with respect to $K$ (see for example \cite{Jackson}). For our problem, $K$ is a frame comoving with the black hole and $K^\star$ is the frame comoving with the qubit.\\
With this analysis of the normalization with respect to $\langle.|.\rangle_{\Gamma(TM,E)}$ we can say that:
\begin{itemize}
\item $\langle \psi|$ is the linear functional associated with the (non-normalized) probability amplitude to find the spin in the state $\psi$ measured by an observer comoving with the black hole ($\langle \psi|\psi\rangle_{\mathbb C^2}=2S^0$).
\item $\langle \psi^\star|$ is the linear functional associated with the (normalized) probability amplitude to find the spin in the state $\psi$ measured by an observer comoving with the qubit ($\langle \psi^\star|\psi\rangle_{\mathbb C^2}=2S^{0\star}=1$).
\end{itemize}  
(with $\|\psi\|_{\Gamma(TM,E)} = 1$).

\subsection{Adiabatic approximation}
We consider the Schr\"odinger like equation for the localized qubit:
\begin{equation}
\imath \frac{d\psi}{d\tau} = H \psi
\end{equation}
with the Hamiltonian $H=H_0 + H_\sharp$ (non-self-adjoint with respect to $\langle.|.\rangle_{\mathbb C^2}$, $H^\dagger \not= H$):
\begin{eqnarray}
H_0 & = & - \omega_\mu^{a0} \dot x^\mu L_{a0} \\
\label{H0} & = & \frac{\imath}{2} \left(\begin{array}{cc} \omega^{03} & \omega^{01} - \imath \omega^{02} \\ \omega^{01} + \imath \omega^{02} & - \omega^{03} \end{array} \right)
\end{eqnarray}
\begin{eqnarray}
H_\sharp & = & - \frac{1}{2} \omega_\mu^{ab} \dot x^\mu L_{ab} \\
& = & -\frac{1}{2} \left(\begin{array}{cc} \omega^{12} & \omega^{23} - \imath \omega^{31} \\ \omega^{23} + \imath \omega^{31} & - \omega^{12} \end{array} \right)
\end{eqnarray}
($\omega^{AB} \equiv \omega^{AB}_\mu \dot x^\mu$). We have $H_0^\dagger = - H_0$ (dissipation operator, see section \ref{origin}) and $H_\sharp^\dagger = H_\sharp$ (Hamiltonian of the qubit rotation). Finally we can write
\begin{equation}
  \label{Hlqubit}
H = \frac{1}{2} \left(\begin{array}{cc} z^3 & z^1 - \imath z^2 \\ z^1 + \imath z^2 & -z^3 \end{array} \right)
\end{equation}
where $z^i = \imath \omega^{0i} - \frac{1}{2} {\epsilon^i}_{jk} \omega^{jk}$ ($z^i_\mu = \imath \omega^{0i}_\mu - \frac{1}{2} {\epsilon^i}_{jk} \omega^{jk}_\mu$ is the complex self-dual Lorentz connection). From the viewpoint of the qubit, the interaction with the gravitational field is similar to a spin submitted to a complexified magnetic field. Let $z=(z^1,z^2,z^3) \in \mathbb C^3$.\\
To integrate the dynamics involved by the Schr\"odinger like equation, we propose to use the adiabatic approximation for the non-self-adjoint Hamiltonians \cite{Nenciu}:
\begin{equation}
\label{adiabtransp}
\psi(\tau) \simeq \sum_{k \in \{+,-\}} \langle \phi_k^*(z(0))|\psi(0)\rangle_{\mathbb C^2} e^{-\imath \int_0^\tau \lambda_k d\tau - \int_{\Gamma} \mathsf A_k} \phi_k(z(\tau))
\end{equation}
where $\phi_k$, $\phi_k^*$ and $\lambda_k$ are respectively the instantaneous right eigenvectors, left eigenvectors and eigenvalues of $H$:
\begin{equation}
H(z) \phi_k(z) = \lambda_k(z) \phi_k(z) \qquad H(z)^\dagger \phi_k^*(z) = \overline{\lambda_k(z)} \phi_k^*(z)
\end{equation}
(the overline denoting the complex conjugation), $\langle \phi_k^*|\phi_q\rangle_{\mathbb C^2} = \delta_{kq}$, and $\mathsf A_k$ are the generators of the non-unitary geometric phases:
\begin{equation}
\mathsf A_k(z) = \langle \phi_k^*(z)|d_{\mathbb C^3}|\phi_k(z) \rangle_{\mathbb C^2}
\end{equation}
$\Gamma$ is the path in $\mathbb C^3$ defined by $\tau \mapsto z(\tau) = (\imath \omega^{0i}_\mu(x(\tau)) \dot x^\mu -\frac{1}{2} {\epsilon^i}_{jk} \omega^{jk}_\mu(x(\tau)) \dot x^\mu)_{i=1,2,3}$ for the geodesic $\tau \mapsto x(\tau)$ followed by the qubit.\\
A simple calculation shows that
\begin{equation}
\lambda_\pm(z) = \pm \frac{1}{2} \sqrt{(z^1)^2+(z^2)^2+(z^3)^2} \equiv \pm \frac{1}{2} \zeta
\end{equation}
\begin{eqnarray}
|\phi_+(z) \rangle & = & \frac{1}{\sqrt{2\zeta(\zeta+z^3)}} \left(\begin{array}{c} \zeta+z^3 \\ z^1+\imath z^2 \end{array} \right) \\
 |\phi_+^*(z) \rangle & = & \frac{1}{\sqrt{2\bar \zeta(\bar \zeta+\overline{z^3})}} \left(\begin{array}{c} \bar \zeta+\overline{z^3} \\ \overline{z^1}+\imath \overline{z^2} \end{array} \right) \\
|\phi_-(z) \rangle & = & \frac{1}{\sqrt{2\zeta(\zeta+z^3)}} \left(\begin{array}{c} - z^1+\imath z^2 \\ \zeta+ z^3 \end{array} \right) \\
 |\phi_-^*(z) \rangle & = & \frac{1}{\sqrt{2\bar \zeta(\bar \zeta+\overline{z^3})}} \left(\begin{array}{c} -\overline{z^1}+\imath \overline{z^2} \\ \bar \zeta + \overline{z^3}\end{array} \right)
\end{eqnarray}
\begin{equation}
\mathsf A_{\pm}(z) = \pm \frac{\imath}{2} \frac{z^1dz^2 - z^2dz^1}{\zeta(\zeta+z^3)}
\end{equation}
The adiabatic approximation is valid if the non-adiabatic coupling is negligible, i.e.
\begin{equation}
\label{valad}
\mathsf N_{-+} = \left|\frac{\langle \phi_-^*(z(\tau)) | \dot H(z(\tau)) | \phi_+(z(\tau)) \rangle_{\mathbb C^2}}{\lambda_+(z(\tau))-\lambda_-(z(\tau))} \right| \ll 1
\end{equation}

Let $\mathcal A$ be the space of the $SL(2,\mathbb C)$-connections of the principal bundle $P$. The eigenvectors can be considered as maps $\hat \phi_\pm : \mathcal A \times TM \to \mathbb C^2$ such that $\hat \phi_\pm(\omega,u) = \phi_\pm(\xi i_u \omega)$, where $i$ is the inner product of $M$ and $\xi : \mathfrak{sl}(2,\mathbb C) \to \mathbb C^3$ is defined by $\xi(\omega^{AB}L_{AB}) = (\imath \omega^{0i} - \frac{1}{2} {\epsilon^i}_{jk} \omega^{jk})_{i=1,2,3}$ ($\{L_{AB}\}_{A,B}$ constitutes a set of generators of $\mathfrak{sl}(2,\mathbb C)$ the Lie algebra of $SL(2,\mathbb C)$). The eigenvectors being defined up to an arbitrary normalization and phase factor, they define $\mathbb C$-line bundles $\mathbf{\Phi}_\pm \to \mathcal A \times TM$ with local trivializations $\tilde \phi_\pm : \mathcal A \times TM \times \mathbb C \xrightarrow{\simeq} \mathbf{\Phi}_\pm$ with $\tilde \phi_\pm(\omega,u,\lambda) = \lambda \hat \phi_\pm(\omega,u)$. $\psi$ obtained by the adiabatic transport formula \ref{adiabtransp} is then a local section of $\mathbf{\Phi}_- \oplus \mathbf{\Phi}_+$ over $\mathcal A \times T\mathcal C$. Note that the left eigenvectors do not define line bundles since their normalization factors are fixed by those of the right eigenvectors.\\
Finally the geometric structure in which the qubit transport takes place can be summarized by the following commutative diagram:
$$
\begin{CD}
E \oplus \bar E @>{P_{(1/2,0)}}>> E @<{\iota_\omega^*}<< \mathbf{\Phi}_- \oplus \mathbf{\Phi}_+ @<{i^* \xi^*}<< \mathbb C\phi_+(\mathbb C^3) \oplus \mathbb C\phi_-(\mathbb C^3) \\
@VVV @VVV @VVV @VVV \\
TM @= TM @>{\iota_\omega}>> \mathcal A \times TM  @>{\xi i_{\Pr_2} \Pr_1}>> \mathbb C^3 \\
@VVV @VVV @VVV \\
M @= M @= M
\end{CD}
$$
where $\iota_\omega(u) = (\omega,u) \in \mathcal A \times TM$. We can note that $\omega \in \mathcal A$ is a connection of the principal bundle $P$ and $\mathsf A_\pm$ are connections of the $\mathbb C^*$-principal bundles associated with $\mathbf{\Phi}_\pm$. We have then three kinds of gauge changes associated with each floor of the composite bundle:
\begin{itemize}
\item ground floor: $\phi \in \mathrm{Diff}M$ (diffeomorphism of the spacetime manifold), $\tilde \omega = \phi^* \omega$ and $\tilde u = \phi_* u$.
\item first floor: $\Lambda \in \mathscr C^\infty(M,SO^+(3,1))$ (local Lorentz transformation), $\tilde \omega^A_{\mu B} = {\Lambda_C}^A \omega^C_{\mu D} {\Lambda^D}_B + {\Lambda_C}^A \partial_\mu {\Lambda^C}_B$ and $\tilde u_A = {\Lambda_A}^B u_B$.
\item second floor: $\mu_\pm \in \mathscr C^\infty(\mathcal A \times TM,\mathbb C^*)$ (normalization and phase local change), $\tilde {\mathsf A}_\pm = \mathsf A_\pm + d\ln \mu_\pm$.
\end{itemize}

The different steps of the construction of the localized qubit adiabatic state can be summarized as follows:
$$ \int_{\mathbb R}^\oplus L^2(N^+\Sigma_\tau,E\oplus \bar E)d\tau \xrightarrow{WKB} \Gamma(T\mathcal C,E\oplus \bar E) \xrightarrow{P_{(\frac{1}{2},0)}} \Gamma_i(T\mathcal C,E) \simeq \Gamma(\mathcal C,E^+) \xrightarrow{adiab.} \Gamma(\mathcal A \times T\mathcal C,\mathbf{\Phi}_- \oplus \mathbf{\Phi}_+) $$

It can be interesting to note that the holonomy of $\omega \in \mathcal A$ along $\mathcal C$ (between $0$ and $\tau$) is
\begin{eqnarray}
\mathrm{Hol}(\omega,\mathcal C) & = & \Pe^{- \imath \int_{\mathcal C} z^i_\mu \sigma_i dx^\mu} \\
& = & \Te^{-\imath \int_0^\tau H(z(\tau))d\tau} \\
& \simeq & \sum_{k\in \{+,-\}} e^{-\imath \int_0^\tau \lambda_k d\tau - \int_\Gamma \mathsf{A}_k} |\phi_k(z(\tau))\rangle\langle \phi_k(z(0))|
\end{eqnarray}
where $\Pe$ and $\Te$ denote path and time ordered exponentials (Dyson series). $\varphi_{\mathcal C,\pm}(\omega) =e^{-\imath \int_0^\tau \lambda_\pm d\tau - \int_\Gamma \mathsf{A}_\pm} $ which characterize the adiabatic state of the qubit can be viewed as cylindrical functions of the space of Lorentz connections, $\varphi_{\mathcal C,\pm} \in \mathrm{Cyl}(\mathcal A)$, and $\psi$ as a linear combination of these two cylindrical functions. It can be interesting to note that (the topological completion of) $\mathrm{Cyl}(\mathcal A)$ constitutes the kinematical Hilbert space of the loop quantum gravity \cite{Rovelli}, this could be indicate a possible connection of the adiabatic localized qubit formalism with a semi-classical limit of the loop quantum gravity.\\
It is also interesting to note that the localized qubit Hamiltonian \ref{Hlqubit} takes the form $H = \frac{1}{2} z^i \sigma_i$ with $\{\sigma_i\}_{i=1,2,3}$ the Pauli matrices and $z^i$ the complex self-dual Lorentz connection. Some D-brane matrix models are governed by an effective Hamiltonian $H^{eff}_{MM} = (Z^i - z^i) \otimes \sigma_i$ where $\{Z^i\}_{i=1,2,3}$ are matrices corresponding to non-commutative coordinates of a stack of D-branes and $z^i$ are scalars corresponding to the position of a probe brane \cite{Berenstein,Badyn}. The eigenequation $H^{eff}_{MM} |\Lambda \rangle = 0$ (with $|\Lambda \rangle \in \mathcal K \otimes \mathbb C^2$ where $\mathcal K$ is a representation Hilbert space for $\{Z^i\}_i$) is associated with the emergent non-commutative geometry of membranes \cite{Badyn} and can be used to study quantum aspects of black holes \cite{Berenstein}. The matrices $Z^i$ can be splitt into a background part and a fluctuation part which is associated to a Lorentz connection \cite{Valtancoli}. Since $H^{eff}_{MM} |\Lambda \rangle = 0 \iff Z^i \otimes \sigma_i |\Lambda \rangle = z^i \sigma_i |\Lambda \rangle$, we see that the localized qubit Hamiltonian \ref{Hlqubit} could be viewed as a kind of a non-commutative eigenvalue of the matrix model. This suggests a possible connection between the localized qubit theory with matrix models and then with supergravity (due to the correspondance between the two theories \cite{Alwis}). Moreover, non-commutative eigenequations as $Z^i \otimes \sigma_i |\Lambda \rangle = z^i \sigma_i |\Lambda \rangle$ appear also in the adiabatic theory of entangled quantum systems and their operator valued geometric phases \cite{Viennot3,Viennot4}. It could be then possible that the connection between the localized qubit theory and D-brane matrix models enlighten the qubit/black-hole correspondence \cite{Levay, Borsten}, where some properties of STU black holes are in correspondence with the entanglement states of several qubits.\\

Note that the evolution governed by $H(z)$ is unitary with respect to $\langle.|.\rangle_{\Gamma(TM,E)}$ (for a proper observer comoving with the qubit) (see \cite{Palmer}): $\langle \psi(\tau)^\star|\psi(\tau) \rangle_{\mathbb C^2} = \langle \psi(0)^\star|\psi(0) \rangle_{\mathbb C^2} \iff \langle \mathrm{Hol(\omega,\mathcal C)} \psi(0)| \bar \sigma^A | \mathrm{Hol(\omega,\mathcal C)} \psi(0) \rangle_{\mathbb C} u_A(\tau) = \langle\psi(0)| \bar \sigma^A |  \psi(0) \rangle_{\mathbb C} u_A(0)$. But it is not unitary with respect to $\langle .|. \rangle_{\mathbb C^2}$ (for an observer comoving with the black hole). We begin to examine this point in the next section.\\

{\it To simplify the notation, from this point we denote $\langle .|.\rangle_{\mathbb C^2}$ only by $\langle .|.\rangle$.}

\subsection{The complex magnetic monopole}
The adiabatic transport of a non-self-adjoint two-level quantum system has been extensively studied in the litterature (see for example \cite{Nesterov, Mostafazadeh, Viennot2, Dridi}). The interesting effects in the adiabatic transport (eq. \ref{adiabtransp}) are related to the submanifold $\mathfrak M$ of $\mathbb C^3$ defined by the crossings $\lambda_+(z) = \lambda_-(z)$. Firstly because the validity of the adiabatic approximation (eq. \ref{valad}) needs to do not approach $\mathfrak M$ (except if $\langle \phi_-^*|\dot H|\phi_+\rangle=0$). Secondly because $\mathfrak M$ is a kind of hybercone separating the region of $\mathbb C^3$ for which $\lambda_\pm$ are real (and generate only pure phases) from the region for which $\lambda_\pm$ are complex (and generate non-unitary (for $\langle.|.\rangle_{\mathbb C^2}$) evolution modifying the relative weights of the superposition of $\phi_\pm$). Since in the self-adjoint case, the geometric phase generator is similar to a magnetic field induced by a magnetic monopole at the eigenvalue crossing point (see \cite{Nakahara}), for the non-unitary case, $\mathfrak M$ has been called complex magnetic monopole \cite{Nesterov} (but $\mathfrak M$ is not an isolated point and is associated with exceptional crossings, i.e. $H(z)$ is not diagonalizable on $\mathfrak M$). Let $\vec \omega^0 = (\omega^{01}, \omega^{02}, \omega^{03})$ and $\vec \omega^\sharp = (\omega^{23},\omega^{31},\omega^{12})$. $\zeta^2 = (\vec \omega^\sharp - \imath \vec \omega^0)^2 = \|\vec \omega^\sharp\|^2 - \|\vec \omega^0\|^2 - 2\imath \vec \omega^\sharp \cdot \vec \omega^0$. Since $\lambda_+(z) = \lambda_-(z) \iff \zeta=0$, the complex magnetic monopole is defined by
\begin{equation}
\mathfrak M : \cases {\|\vec \omega^0 \| = \|\vec \omega^\sharp\| \\ \vec \omega^0 \cdot \vec \omega^\sharp = 0}
\end{equation}
$\dim_{\mathbb R} \mathfrak M = 4$. If the condition $\vec \omega^0 \cdot \vec \omega^\sharp = 0$ is satisfied, outside $\mathfrak M$ ($\|\vec \omega^\sharp\| > \|\vec \omega^0\|$), $\zeta \in \mathbb R$ and $e^{-\imath \int \lambda_\pm d\tau} \in U(1)$ are just pure phases; but inside $\mathfrak M$ ($\|\vec \omega^\sharp\| < \|\vec \omega^0\|$), $\zeta \in \imath \mathbb R$ and $e^{-\imath \int \lambda_\pm d\tau} \in \mathbb R^+$ are non-unitary dynamical phases. In this last case, the evolution modifies the weights of the superposition of $\phi_\pm$ (with respect to $\langle .|. \rangle_{\mathbb C^2}$, i.e. for an observer comoving with the black hole). We will call this effect a \textbf{dynamical decoherence}, because the following coherence 
\begin{eqnarray}
\frac{|\langle \phi_+^*|\psi\rangle \langle \psi|\phi_-^* \rangle|}{\|\psi\|^2} & = & \frac{|c_+c_-| e^{\frac{1}{2} \int_0^\tau |\zeta| d\tau} e^{- \frac{1}{2} \int_0^\tau |\zeta| d\tau}}{|c_+|^2 e^{\int_0^\tau |\zeta| d\tau} + |c_-|^2 e^{-\int_0^\tau |\zeta| d\tau}} \\
& \sim & \left|\frac{c_-}{c_+}\right| e^{- \int_0^\tau |\zeta|d\tau}
\end{eqnarray}
falls to zero for large $\tau$ (with $c_k = \langle \phi_k^*(z(0))|\psi(0)\rangle$, we have neglected the geometric phases and supposed that $\Im \zeta=|\zeta|>0$). We have not considered the effects of the non-unitary geometric phases $e^{- \int_\Gamma \mathsf{A}_\pm}$ which can induce a \textbf{geometric decoherence} if $\mathsf{A}_\pm \in \mathbb R$ (we call it geometric decoherence since the geometric phase depends only on the shape of the followed path $\Gamma$ and not from the proper time).\\
Let $\mathfrak M_{\omega} = \imath_{\omega}^{-1} \xi^{-1} (\mathfrak M)$ be the preimage of $\mathfrak M$ into $TM$. It is important to note that the complex magnetic monopole for a fixed spacetime geometry $\mathfrak M_{\omega}$ is not a submanifold of the spacetime $M$ but a submanifold of the tangent bundle $TM$. The complex magnetic monopole around the black hole ``viewed'' by the qubit depends on its four-velocity. Gobally the set of all complex magnetic monopoles is $\mathfrak M_{\mathcal A} = \{(\omega,\mathfrak M_{\omega}),\omega \in \mathcal A\} \subset \mathcal A \times TM$. In some cases, a class $\mathscr G_{\{I_\alpha\}_\alpha}$ of geodesics can be defined with some first integrals $\{I_\alpha\}_\alpha$ and $i_u \omega$ ($u\in T_x \mathcal C$, $\mathcal C \in \mathscr G_{\{I_\alpha\}_\alpha}$) depends only on $\{I_\alpha\}_\alpha$ and $x$. In that case $\mathfrak M_{\omega,\{I_\alpha\}_\alpha} = \pi_T\left(\mathfrak M_\omega \cap \pi_T^{-1}(\mathscr G_{\{I_\alpha\}_\alpha}) \right)$ is a submanifold of $M$ which is an image (for the qubits following geodesics of $\mathscr G_{\{I_\alpha\}_\alpha}$) of the complex magnetic monopole in the spacetime.

\subsection{Physical origin of the non-unitarity evolution}
\label{origin}
To understand the physical origin of the non-unitarity with respect to $\langle .|.\rangle_{\mathbb C^2}$ (observer comoving with the black hole), consider first a more simple model constituted by a three-level system (with levels denoted by $\{|d\rangle,|0\rangle,|1\rangle\}$), governed by an Hamiltonian $H$ and with spontaneous emission from $|0\rangle$ to the ``dark state'' $|d\rangle$ with a rate $\gamma_-$. We want to consider the system restricted to $(|0\rangle,|1\rangle$) as a qubit and to forget the dark state $|d\rangle$. The system obeys to a master equation \cite{Breuer}:
\begin{equation}
\frac{d\rho}{dt} = -\imath[H,\rho] -\frac{\gamma_-}{2} \{\sigma_{d0}^+ \sigma_{d0}^- ,\rho\} + \gamma_- \sigma^-_{d0} \rho \sigma^+_{d0}
\end{equation}
where $\rho$ is the density matrix of the system, $\{.,.\}$ denotes the anti-commutator, $\sigma_{d0}^- = |d\rangle\langle 0|$ and $\sigma_{d0}^+ = |0\rangle\langle d|$. The equation can be rewritten as
\begin{equation}
\label{mastereq}
\frac{d\rho}{dt} = -\imath(H^{eff}\rho - \rho H^{eff \dagger}) + \gamma_- \rho_{00} |d\rangle \langle d|
\end{equation}
where $H^{eff} = H - \imath \frac{\gamma_-}{2} |0\rangle \langle 0|$. The anti-self-adjoint part of the effective Hamiltonian $- \imath \frac{\gamma_-}{2} |0\rangle \langle 0|$ models the lost of population from $|0\rangle$ to the dark state by spontaneous emission, whereas $\gamma_- \rho_{00} |d\rangle \langle d|$ models the gain of population of this dark state. So, if we forget the dark state in the modelization, the qubit obeys to a Schr\"odinger equation governed by a non-self-adjoint effective Hamiltonian $H^{eff}_{|(|0\rangle,|1\rangle)} = H_{|(|0\rangle,|1\rangle)} -  \imath \frac{\gamma_-}{2} |0\rangle \langle 0|$. The non-self-adjoint part of $H^{eff}$, $-  \imath \frac{\gamma_-}{2} |0\rangle \langle 0|$, can be called a dissipation operator, since it models the dissipation of the wave function induced by the lost of population from $|0\rangle$ to the dark state. Dynamically, it generates a factor $e^{- \gamma_- t}$ on the population $|\langle 0|\psi(t)\rangle|^2$ (for $\psi$ a solution of the Schr\"odinger equation governed by $H^{eff}$) killing it with the time (reproducing the relaxation described by the master equation, which induces the fall of the population of $|0\rangle$). More precisely, if direct couplings with the dark state do not occur, i.e. $\langle 0|H|d\rangle = \langle 1|H|d\rangle = 0$, then the populations and coherences of the two active states obeys to
\begin{equation}
\label{popcohe}
\frac{d\rho_{ij}}{dt} = -\imath\left(H^{eff}_{i0} \rho_{0j} + H^{eff}_{i1} \rho_{1j} - \overline H^{eff}_{j0} \rho_{i0} - \overline H^{eff}_{j1} \rho_{i1} \right)
\end{equation}
$\forall i,j \in \{0,1\}$. Let $\psi$ be the solution of $\imath \dot \psi = H^{eff}_{|(|0\rangle,|1\rangle)} \psi$, then $P = |\psi \rangle \langle \psi|$ obeys to $\dot P =- \imath\left( H^{eff}_{|(|0\rangle,|1\rangle)} P - P H^{eff \dagger}_{|(|0\rangle,|1\rangle)} \right)$ and the populations and the coherences $P_{ij} = \langle i |P|j \rangle$ obeys to the same equation \ref{popcohe}. It follows that $\rho_{ij} = P_{ij}$ (note that $P^2 \not= P$ since $\psi$ is not normalized due to the non-hermitian character of $H^{eff}_{|(|0\rangle,|1\rangle)}$). We see that if the dark state is not directly coupled with the active states (except by the spontaneous emission), the non-hermitian hamiltonian $H^{eff}_{|(|0\rangle,|1\rangle)}$ generates for the active states the same evolution than the master equation.\\
In a curved spacetime there is an ambiguity concerning the particle number. Due to the Unruh effect, the vaccum in the rest frame becomes a thermal state in an uniformly accelerated frame \cite{Fuentes,Fuentes2}. At the level of the quantum field theory in curved spacetime, the evolution in the black hole frame of the Dirac field spontaneously couples the one particle state to the zero particle state (in the fermionic Fock space of the system). But the semi-classical and WKB approximations used in our model forgets this last one (we want to have one and only one qubit). We have then only two qubit states $|1_0\rangle$ and $|1_1\rangle$ (forming the canonical basis of $\mathbb C^2$ used in the construction of the different bundles), and a dark state: the vacuum $|\varnothing\rangle$. In the same way that for the small example of a three-level system, the qubit is then governed by a non-self-adjoint effective Hamiltonian. We will study this point with more details in a concrete example in section 4.\\
We call the non-self-adjoint part of the Hamiltonian, $H_0$ (equation \ref{H0}), a dissipation operator in the sense where it describes a relaxation phenomenon as in the example of a three level atom with a dark state. We find in the literature a lot of examples of physical systems which can be described by non-hermitian Hamiltonians in order to model a relaxation process by a dissipation behaviour and which are in accordance with experimental studies. We can cite for example the modelling of a spontaneous decay \cite{Lee1}, of a finite lifetime state \cite{Lee2} or of a quantum resonance in atomic or molecular systems \cite{Moiseyev}.

\section{Quantum teleportation}
Let Alice and Bob be initially at the point $x_{\mathbf{B}}$ of $M$, supposed sufficiently far from the black hole to consider that $M$ is flat in the neighbourhood of $x_{\mathbf{B}}$. We set $|0\rangle = \left(0 \atop 1\right)$ and $|1\rangle = \left(1 \atop 0\right)$. Bob is supposed comoving with the black hole (he stays at $x_{\mathbf B}$), but Alice follows a geodesic going near the event horizon at a point $x_{\mathbf A}$. Alice wants to teleport information when she will be at this point. At the moment $\tau_{\mathbf A} = \tau_{\mathbf B} = 0$ when Alice leaves Bob, they have an entangled qubit pair in a Bell state:
\begin{equation}
|\psi_{\A \B}^0 \rrangle = \frac{1}{\sqrt{2}} (|0_\A 0_\B \rrangle + |1_\A 1_\B \rrangle) \in \pi_E^{-1}(u_\A^0) \otimes \pi_E^{-1}(u_\B^0)
\end{equation}
$u_\A^0 \in T_{x_\B}M$ and $u_\B^0 = (1,0,0,0) \in T_{x_\B}M$ are the initial four-velocities of Alice and Bob. Since Alice and Bob belongs to two different frames, each one has its proper definition of the qubit states. For Bob, the linear functionals of finding its qubit in a particular state are $\langle 0|$ and $\langle 1|$, involving that $|0_\B\rangle = |0\rangle$ and $|1_\B\rangle = |1\rangle$. But for Alice, her linear functionals are $\langle 0^\star|$ and $\langle 1^\star|$ since she is not comoving with the black hole. The qubit states for Alice are then defined by $\langle a^\star|b_\A \rangle = \langle a|\bar \sigma^A|b_\A\rangle u^0_{\A A} = \delta_{ab}$ ($\forall a,b \in\{0,1\}$). It follows that $|0_\A \rangle = \sigma^A u^0_{\A A} |0\rangle$ and $|1_\A \rangle = \sigma^A u^0_{\A A} |1\rangle$ ($(\bar \sigma^A u^0_{\A A})^{-1} = \sigma^A u^0_{\A A}$).
\begin{equation}
|\psi^0_{\A\B} \rrangle = \frac{1}{\sqrt{2}} \sum_{ab} \langle a|\sigma^A|b\rangle u_{\A A}^0 |ab \rrangle
\end{equation}
In the flat region ($z_\B = 0$), we have $|\phi_+(z_\B)\rangle = \frac{1}{\sqrt 2}\left(1 \atop 1 \right)$ and $|\phi_-(z_\B)\rangle = \frac{1}{\sqrt 2}\left(-1 \atop 1 \right)$.
\begin{equation}
|\psi^0_{\A\B} \rrangle = \frac{1}{2} \sum_{a,b=0}^1 \sum_{i=\pm} \langle a|\sigma^A|b\rangle u_{\A A}^0 i^a |\phi_i(z_\B) \rangle \otimes |b \rangle
\end{equation}
Let $\tau_\A^1$ the proper time when Alice arrives at $x_\A$. We suppose that the evolution along the geodesic $\mathcal C$ linking $x_\B$ to $x_\A$ is adiabatic for the Alice's qubit (as eq. \ref{adiabtransp}). We have then for $\tau_\A = \tau_\A^1$ and $\tau_\B>0$
\begin{equation}
|\psi^1_{\A\B} \rrangle = \frac{1}{2} \sum_{abi} \langle a|\sigma^A|b\rangle u_{\A A}^0 i^a e^{\imath \varphi_i} |\phi_i(z_\A) \rangle \otimes |b \rangle
\end{equation}
($\varphi_i = - \int_0^{\tau_\A^1} \lambda_i d\tau + \imath \int_\Gamma \mathsf A_i$). Note that $|\psi^1_{\A\B} \rrangle$ is defined for two proper times, one for Alice and one for Bob, since their clocks are asynchronous. The evolution for the Bob's qubit is trivial since it is inertial in a flat part of $M$.
\begin{equation}
|\phi_i(z_\A)\rangle = \sum_c \langle c^\star|\phi_i(z_\A)\rangle |c_\A\rangle = \sum_c \langle c|\bar \sigma^C|\phi_i(z_\A) \rangle u_{\A C}^1 |c_\A \rangle
\end{equation}
where $u_\A^1 \in T_{x_\A}M$ is the Alice's four-velocity at $\tau_\A^1$.
\begin{equation}
|\psi_{\A\B}^1 \rrangle = \frac{1}{\sqrt 2} \sum_{bc} \chi_{bc} |c_\A b \rrangle
\end{equation}
with
\begin{equation}
\chi_{bc} = \frac{1}{\sqrt 2} \sum_{ai} \langle a|\sigma^A|b\rangle u^0_{\mathbf A A} i^a e^{\imath \varphi_i} \langle c|\bar \sigma^C|\phi_i(z_\A)\rangle u^1_{\A C}
\end{equation}
Alice encodes a quantum information in a qubit $|\psi_I \rangle = \alpha |0_\A \rangle + \beta |1_\A \rangle$ ($|\alpha|^2+|\beta|^2=1$). The state of the three qubits is then $|\psi^1_{\A\A\B} \rrrangle = |\psi_I \rangle \otimes |\psi^1_{\A\B} \rrangle$. Alice performs then the operations of the usual teleportation protocol:
\begin{equation}
|\psi^2_{\A\A\B} \rrrangle = ({\mathsf H}_\A \otimes \id \otimes \id) (\mathsf{CNOT}_\A \otimes \id) |\psi^1_{\A\A\B}\rrrangle
\end{equation}
where $\mathsf{CNOT}_\A$ and $\mathsf H_\A$ are the CNOT and Hadamard gates in the Alice's frame. After some algebra, we find
\begin{eqnarray}
|\psi^2_{\A\A\B} \rrrangle & = & |0_\A 0_\A \rrangle \otimes\left(\frac{\alpha \chi_{00}+\beta \chi_{10}}{2} |0\rangle + \frac{\alpha\chi_{01}+\beta \chi_{11}}{2} |1 \rangle \right) \nonumber \\
& & + |1_\A 0_\A \rrangle \otimes\left(\frac{\alpha \chi_{00}-\beta \chi_{10}}{2} |0\rangle + \frac{\alpha\chi_{01}-\beta \chi_{11}}{2} |1 \rangle \right) \nonumber \\
& & + |0_\A 1_\A \rrangle \otimes\left(\frac{\alpha \chi_{10}+\beta \chi_{00}}{2} |0\rangle + \frac{\alpha\chi_{11}+\beta \chi_{01}}{2} |1 \rangle \right) \nonumber \\
& & + |1_\A 1_\A \rrangle \otimes\left(\frac{\alpha \chi_{10}-\beta \chi_{00}}{2} |0\rangle + \frac{\alpha\chi_{11}-\beta \chi_{01}}{2} |1 \rangle \right) 
\end{eqnarray}
Alice performs a measurement of her qubits. To fix the discussion, we suppose that she finds $0_\A 0_\A$ (the result can be easily adapted for another result). Alice sends to Bob by a classical communication chanel what is the operation to perform on his qubit (in our example, the operation is the identity). Bob receives the message at $\tau_\B^3$. The state is then for $\tau_\A > \tau_\A^1$ and $\tau_\B = \tau_\B^3$:
\begin{eqnarray}
|\psi^3_{\A\A\B}\rrrangle & = & \left(U_\A\otimes U_\A |0_\A 0_\A \rrangle\right) \nonumber \\
& & \qquad \otimes \left((\alpha \chi_{00}+\beta \chi_{10}) |0\rangle + (\alpha\chi_{01}+\beta \chi_{11}) |1 \rangle \right)
\end{eqnarray}
where $U_\A$ is the evolution operator for an Alice's qubit after $\tau_\A^1$. The fidelity of the quantum teleportation is then
\begin{eqnarray}
\label{fidelity}
F(\alpha,\beta) & = & \frac{|(\bar \alpha \langle 0| + \bar \beta \langle 1| ) ((\alpha \chi_{00}+\beta \chi_{10}) |0\rangle + (\alpha\chi_{01}+\beta \chi_{11}) |1 \rangle)|}{\|(\alpha \chi_{00}+\beta \chi_{10}) |0\rangle + (\alpha\chi_{01}+\beta \chi_{11}) |1 \rangle\|} \\
& = & \frac{\left||\alpha|^2 \chi_{00} + \bar \alpha \beta \chi_{10} + \alpha \bar \beta \chi_{01} + |\beta|^2 \chi_{11} \right|}{\sqrt{|\alpha \chi_{00} +\beta \chi_{10}|^2+|\alpha\chi_{01}+\beta \chi_{11}|^2}}
\end{eqnarray}
The fidelity of the teleportation is clearly degraded by the decoherence induced by the black hole which is encoded in $\chi_{bc}$.\\
Remark: for a flat spacetime with Alice having a constant four-velocity we have $\chi_{bc} = \sum_a \langle c|\bar \sigma^C|a\rangle \langle a|\sigma^A |b \rangle u_{\A A} u_{\A C} = \langle c | \bar \sigma^C u_{\A C} \sigma^A u_{\A A} |b \rangle = \delta_{cb}$, and then $F=1$ (we refind the efficiency of the usual teleporation protocol).

\section{Applications}
\subsection{Rindler spacetime}
In order to compare with the Fuentes-Schuller Mann model \cite{Fuentes}, we first consider the case of the Rindler spacetime defined by the metric
\begin{equation}
d\tau^2 = (Ax)^2 dt^2 - dx^2
\end{equation}
which corresponds to a flat spacetime viewed in a noninterial frame uniformly accelerated (with acceleration parameter $\frac{1}{A}$), or to the surface gravity approximation of a Schwarzschid black hole ($x=2 \sqrt{r_S(r-r_S)}$ and $A = \frac{1}{2 r_S}$, $r_S = 2GM$ being the Schwarzschild radius) (see \cite{Fuentes}). The tetrad fields are $e^0 = Ax dt$ and $e^1=dx$, and the only one non-zero Lorentz connection component is $\omega^{01} = -Adt$. It follows that $z^1 = -\imath A \dot t$ and $z^2=z^3=0$, and then
\begin{equation}
  \label{HRindler}
H = - \frac{\imath A \dot t}{2} \left(\begin{array}{cc} 0 & 1 \\ 1 & 0 \end{array} \right)
\end{equation}
with $\lambda_\pm = \mp \frac{\imath}{2} A \dot t$ and
\begin{equation}
|\phi_-\rangle = \frac{1}{\sqrt{2}} \left(\begin{array}{c} -1 \\ 1 \end{array} \right), \qquad |\phi_+\rangle = \frac{1}{\sqrt{2}} \left(\begin{array}{c} 1 \\ 1 \end{array} \right) \quad 
\end{equation}
$\mathsf A_\pm = \mathsf N_{-+} = 0$. It is interesting to note that since the qubit moves in the $x$-direction, $\phi_+$ corresponds to a spin parallel to the linear momentum (so to a positive helicity state) and $\phi_-$ to a spin antiparallel to the linear momentum (a negative helicity state) ($\sigma_x \phi_\pm = \pm \phi_\pm$). The geodesic equations are
\begin{equation}
\cases{ \ddot t + \frac{2}{x} \dot t \dot x = 0 \\ \ddot x + A^2 x \dot t^2 = 0 }
\end{equation}
The first geodesic equation defines the first integral:
\begin{equation}
x^2 \dot t = K
\end{equation}
The second geodesic equation becomes the autonomous equation:
\begin{equation}
\label{geodesicRindler}
\ddot x + \frac{A^2K^2}{x^3} = 0
\end{equation}
The adiabatic transport of a qubit state $\psi(\tau_0) = c_+ \phi_+ + c_- \phi_-$ ($c_\pm \in \mathbb C$) is then
\begin{eqnarray}
\label{transpRindler}
\psi(\tau) & = & \frac{c_-}{\sqrt 2} e^{\frac{A}{2}(t(\tau)-t(0))} \left(\begin{array}{c} -1 \\ 1 \end{array} \right) + \frac{c_+}{\sqrt 2} e^{-\frac{A}{2}(t(\tau)-t(0))} \left(\begin{array}{c} 1 \\ 1 \end{array} \right) \\
& = & \frac{c_-}{\sqrt 2} e^{\frac{AK}{2} \int_{0}^\tau \frac{d\tau}{x^2}} \left(\begin{array}{c} -1 \\ 1 \end{array} \right) + \frac{c_+}{\sqrt 2} e^{-\frac{AK}{2} \int_{0}^\tau \frac{d\tau}{x^2}} \left(\begin{array}{c} 1 \\ 1 \end{array} \right)
\end{eqnarray}
The dynamical decoherence kills the positive helicity state in favor of the negative helicity state. We can heuristically understand this fact as follows. The vacuum of the Minkowsky spacetime becomes in the noninertial frame $|\varnothing\rangle = \cos \theta |\varnothing\rangle_I |\varnothing \rangle_{II} + \sin \theta |1_{k,s}\rangle_I |1_{-k,s} \rangle_{II}$ (see \cite{Fuentes2}) where $\tan \theta = e^{-\pi \omega A}$ ($k$ denotes the momentum, $s$ the helicity, $I$ and $II$ denote the two regions separated by the horizon, $\omega=m\dot t$). It is then associated with a density matrix $\rho^{\varnothing}_{k,s} = \tr_{II} |\varnothing\rangle\langle \varnothing| = \cos^2 \theta |\varnothing\rangle \langle \varnothing|_I + \sin^2 \theta |1_{k,s}\rangle \langle 1_{k,s}|_{I}$ which is a thermal distribution with temperature $T = \frac{a}{2\pi k_B}$ ($k_B$ is the Boltzmann constant), corresponding to the Unruh radiation. It follows that the fermion is coupled with this thermal bath. For the positive helicity mode (which is the part of the Weyl spinor with positive energy), we can write that the density matrix $\rho_+$ (for $|1_{k,+}\rangle = |\phi_+\rangle$ and $|\varnothing\rangle$) obeys to the master equation (see for example \cite{Ghosh}):
\begin{eqnarray}
\frac{d\rho_+}{d\tau} & = & - \frac{\gamma}{2} (1-\bar n) \{c_+^+ c_+,\rho_+\} + \gamma(1-\bar n) c_+ \rho_+ c_+^+ \nonumber \\
& & - \frac{\gamma}{2} \bar n \{c_+ c_+^+,\rho_+\} + \gamma \bar n c_+^+ \rho_+ c_+
\end{eqnarray}
where $c_+^\pm$ are the fermionic creation/annihilation operators on the positive helicity mode ($c_+ = |\varnothing\rangle \langle 1_{k,+}|$ and $c_+^+ = |1_{k,+}\rangle \langle \varnothing|$), $\bar n = \frac{1}{e^{\frac{\omega}{k_B T}} + 1}$ and $\gamma$ characterizes the spectral density of the bath. With some assumptions $\gamma(1-\bar n) \simeq \frac{\gamma_0 \omega}{4 k_B T}$ where $\gamma_0$ is a constant (see \cite{Karmakar}). Since $\bar n$ is very small (the Unruh temperature $T$ is very small), the master equation is dominated by the dissipation of $|1_{k,+}\rangle = |\phi_+\rangle$ in accordance with eq. \ref{transpRindler}. By following the approximation explained section \ref{origin} (by projecting onto $|1_{k,+}\rangle \langle 1_{k,+}|$, by neglecting the quantum jumps and the $\bar n$ terms) we have
\begin{equation}
H^{eff}_+ = - \frac{\imath \gamma_0 \omega}{4 k_B T} |\phi_+\rangle \langle \phi_+|
\end{equation}
We refind $H^{eff}_+ = \lambda_+ |\phi_+\rangle \langle \phi_+|$ by setting $\gamma_0 = \frac{1}{\pi m}$. For the negative helicity mode the problem is quite different since it is associated with the part of the Weyl spinor with negative energy. It follows that the roles of $\gamma (1-\bar n)$ and $\gamma \bar n$ are inverted in the master equation. This one is then dominated by the increase of the population of $|1_{k,-}\rangle = |\phi_-\rangle$. We can then postulate an effective Hamiltonian creating negative helicity population, as $H^{eff}_- = \lambda_- |\phi_-\rangle \langle \phi_-|$.\\
This is just an heuristic argument to relate the results obtained with the Fuentes-Schuller Mann model in \cite{Fuentes2} (in curved spacetime quantum field theory with the simple geometry of the Rindler space) with the model of localized qubit (non-hermitian quantum mechanics associated with a curved spacetime). The correct derivation of the non-hermitian Hamiltonian (\ref{HRindler}) follows the method exposed in \cite{Palmer}. The role of this heuristic argument consists to make an analogy between the Unruh effect of the Fuentes-Schuller Mann model which is responsible of the decoherence and the relaxation (because of the entanglement between the qubits separated by the horizon), with the non-hermitian dynamics of the localized qubit model which is responsible of the decoherence and the relaxation of the localized qubit. The two phenomemons (Unruh effect and non-hermitian evolution) model the same thing, the coupling of the qubit with the gravity encoded by the Dirac-Einstein equation \ref{DiracEinstein} (which is the primary equation of the two approaches, Fuentes-Schuller Mann and localized qubit, but treated with different approximations), and have the same consequences (decoherence and relaxation of the qubit state).\\

The geodesic equation (eq. \ref{geodesicRindler}) has for solution $x(\tau) = \sqrt{A^2K^2(\beta+\tau)^2-1}$ with $\beta=-\frac{\sqrt{x(0)^2+1}}{AK}$. We have then an analytical expression of the dynamical phases:
\begin{equation}
e^{\frac{AK}{2} \int_{0}^\tau \frac{d\tau}{x^2}} = \left(\frac{\left(1+AK\beta \right)\left(1-AK(\beta+\tau)\right)}{\left(1-AK\beta \right)\left(1+AK(\beta+\tau)\right)} \right)^{1/4}
\end{equation}
Let $\tau_H = -\frac{1}{AK} - \beta$ be the proper time for which the qubit reaches the horizon. We have $\lim_{\tau \to \tau_H} e^{\frac{AK}{2} \int_{0}^\tau \frac{d\tau}{x^2}} = 0$.\\

\begin{figure}
\begin{center}
\includegraphics[width=8cm]{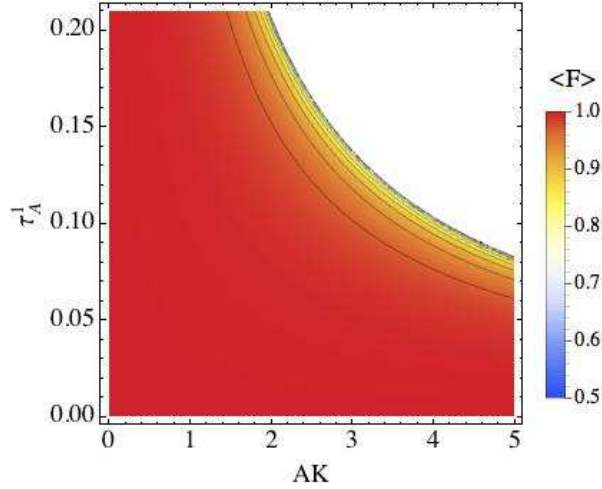}
\caption{\label{fidelity_rindler} Average fidelity $\langle F \rangle = \int_0^{\pi} \int_0^{2\pi} F(\cos(\alpha),e^{\imath \beta}\sin(\alpha)) \frac{d\alpha d\beta}{2\pi^2}$ (eq. \ref{fidelity}) of the teleportation protocol, for Alice following a geodesic of the Rindler spacetime and Bob being static, with respect to $AK$ (first integral of the geodesic) and $\tau^1_\A$ the Alice's proper time when she realizes its part of the protocol. For each value of $AK$, the fidelity is drawn until the proper time when Alice reaches the Rindler horizon.}
\end{center}
\end{figure}

The fidelity of the teleportation protocol for the Rindler spacetime is represented fig. \ref{fidelity_rindler}.
The fidelity of the quantum teleportation falls if Alice approaches too close to the Rindler horizon.

\subsection{Schwarzschild black hole}
We consider the metric associated with a static black hole with spherical symmetry:
\begin{equation}
\label{metricSchwarz}
d\tau^2 = T(r)^2 dt^2 - R(r)^{-2} dr^2 - r^2(d\theta^2+\sin^2\theta d\varphi^2)
\end{equation}
where $T$ and $R$ are the factors of time dilation and length contraction, the Schwarzschild metric being obtained for $T(r)=R(r)= \sqrt{1- \frac{r_S}{r}}$ with $r_S=2GM$ the Schwarzschild radius.  The tetrad fields are $e^0 = T(r)dt$, $e^1=R(r)^{-1}dr$, $e^2=rd\theta$ and $e^3=r \sin \theta d\varphi$; and the non-zero components of the Lorentz connection are $\omega^{01} = - T'(r)R(r)dt$, $\omega^{12}=-R(r)d\theta$, $\omega^{13}=R(r)\sin\theta d\varphi$ and $\omega^{23}=\cos \theta d\varphi$. Because of the spherical symmetry, we can restric our attention to the equatorial plane $\theta=\frac{\pi}{2}$. The geodesic equations are
\begin{equation}
\label{geodesicSchwarz}
\cases{\ddot t + 2 \frac{T'}{T} \dot t \dot r = 0 \\ \ddot r + T'R^2T\dot t^2 - \frac{R'}{R}\dot r^2 -R^2r \dot \varphi^2 = 0 \\ \ddot \varphi + \frac{2}{r} \dot r \dot \varphi = 0}
\end{equation}
The first and the last geodesic equations define the first integrals:
\begin{eqnarray}
T^2 \dot t & = & E \\
r^2 \dot \varphi & = & L
\end{eqnarray}
$E$ and $L$ being the energy and the angular momentum by mass unit. We have $z^1 = -\imath \frac{T'R}{T^2} E$, $z^2=\frac{R}{r^2} L$, and $z^3=0$. It follows that
\begin{equation}
H = \frac{1}{2} \left(\begin{array}{cc} 0 & -\imath \frac{T'R}{T^2} E - \imath \frac{R}{r^2} L \\ -\imath \frac{T'R}{T^2} E + \imath \frac{R}{r^2} L & 0 \end{array} \right)
\end{equation}
with $\lambda_\pm = \pm \frac{1}{2} \sqrt{\frac{R^2}{r^4}L^2-\frac{T^{\prime 2}R^2}{T^4}E^2}$. $\lambda_\pm \in \mathbb R$ if $L \geq \frac{T' r^2}{T^2} E$ ($\mathfrak M_{\omega,L,E}= \left\{(r_{LE},\varphi); \varphi \in [0,2\pi] \text{ with } \frac{T'(r_{LE}) r^2_{LE}}{T(r_{LE})^2} = \frac{L}{E}\right\}$). For the Schwarzschild case, the dynamical decoherence disapears for $\left(1-\frac{r_S}{r}\right)^{3/2} L \geq \frac{r_S}{2} E$, i.e. if $\frac{r_SE}{2L} <1$ and $r > r_{LE} = \frac{r_S}{1-\left(\frac{r_SE}{2L}\right)^{2/3}}$. It follows that the qubit is submitted to dynamical decoherence except if it follows a strongly rotating geodesic ($L$ large) and far from the complex magnetic monopole (which is a sphere of radius $r_{LE} \geq r_S$). The generators of the geometric phases are $\mathsf A_\pm = \pm\frac{1}{2} \frac{uvw'-u'vw-uv'w}{w^2L^2-u^2v^2E^2} ELdr \in \Omega^1(M,\mathbb R)$ (with $u=\frac{T'}{T}$, $v=\frac{R}{T}$ and $w=\frac{R}{r^2}$). For the Schwarzschild case, we have $\mathsf A_\pm = \pm \frac{EL}{2} \frac{r^2_S}{(1-\frac{r_S}{r})^3 r^2 L^2-\frac{r_S^2r^2}{4(1-\frac{r_S}{r})}E^2} dr$. Geometric decoherence is always present except for the radial geodesics ($L=0$) and the circular orbits ($r$ constant). Moreover the non-adiabatic coupling is $\mathsf N_{-+} = \frac{|(wL-uE)(u'w-uw')|}{(w^2L^2-u^2E^2)^{3/2}} LE |\dot r|$, assuring without any assumption concerning the velocity, the validity of the adiabatic approximation for the radial geodesics ($L=0$) and the circular orbits ($\dot r = 0$).

\subsubsection{Radial geodesics for the Schwarzschild metric:}
\begin{figure}
\begin{center}
\includegraphics[width=8cm]{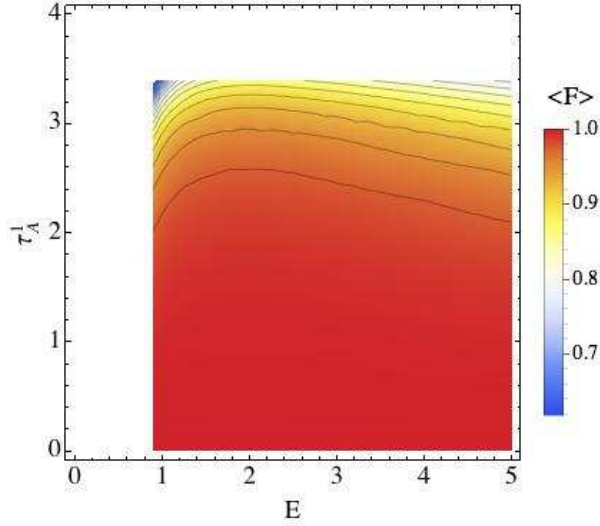}
\caption{\label{fidelity_schwarz_radial} Average fidelity $\langle F \rangle= \int_0^{\pi} \int_0^{2\pi} F(\cos(\alpha),e^{\imath \beta}\sin(\alpha)) \frac{d\alpha d\beta}{2\pi^2}$ (eq. \ref{fidelity}) of the teleportation protocol, for Alice following a radial geodesic of the Schwarzschild spacetime and Bob comoving with the black hole, with respect to $E$ (first integral of the geodesic) and $\tau^1_\A$ the Alice's propre time when she realizes its part of the protocol. The fidelity is drawn until the proper time when Alice reaches the event horizon.}
\end{center}
\end{figure}

We consider first the radial geodesics $L=0$ ($\dot \varphi = 0$). The second geodesic equation (eq. \ref{geodesicSchwarz}) is then reduced to $\ddot r + \frac{r_S}{2r^2} = 0$, which has for solution: $r(\tau)=\left(-3 \sqrt{r_S} \tau + r_0^{3/2}\right)^{2/3}$. The event horizon is reached at $\tau_H=\frac{r_0^{3/2}-r_S^{3/2}}{3 \sqrt{r_S}}$. The fidelity of the teleportation protocol for this situation is drawn figure \ref{fidelity_schwarz_radial}.
As for the Rindler spacetime, the fidelity of the quantum teleportation falls if Alice approaches too close to the event horizon, because of the decoherence induced by the gravitational field (all the radial geodesics are inside $\mathfrak M_{\omega,L=0,E} = \{(+\infty,\varphi), \varphi \in [0,2\pi]\}$).

\subsubsection{Circular orbits for the Schwarzschild metric:}
\begin{figure}
\begin{center}
\includegraphics[width=8cm]{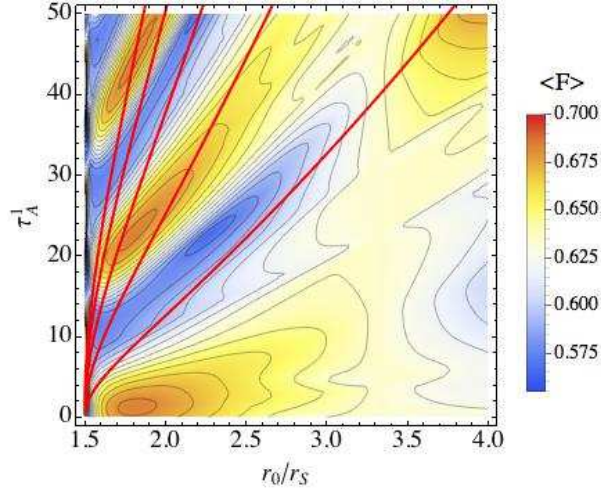}
\caption{\label{fidelity_schwarz_circular} Average fidelity $\langle F \rangle = \int_0^{\pi} \int_0^{2\pi} F(\cos(\alpha),e^{\imath \beta}\sin(\alpha)) \frac{d\alpha d\beta}{2\pi^2}$ (eq. \ref{fidelity}) of the teleportation protocol, for Alice following a circular orbit around the black hole and Bob comoving with the black hole, with respect to $r_0$ (radius of the orbit) and $\tau^1_\A$ the Alice's propre time when she realizes its part of the protocol. The red lines indicates the proper times corresponding to the orbital periods. We start at $r_0=\frac{3}{2}r_S$ (the photon sphere) since no closed orbit exists under this value.}
\end{center}
\end{figure}

We consider the circular orbits defined by $r=r_0$ (constant) and $\varphi(\tau) = \frac{L}{r_0^2} \tau + \varphi_0$. The second geodesic equation (eq. \ref{geodesicSchwarz}) involves that $L^2 = \frac{r_Sr_0^2}{2r_0-3r_S}$ and the metric (eq. \ref{metricSchwarz}) involves that $E^2=T(r_0)^2\left(1+\frac{L^2}{r^2_0}\right)$. The fidelity of the teleportation protocol for this situation is drawn figure \ref{fidelity_schwarz_circular}.
For the circular orbits, the effect is essentially caused by the difference of four-velocity between Alice and Bob (explaining why the fidelity is almost uniform with respect to $r_0$ and $\tau_\A^1$). No decoherence occurs since for all $r_0$, $\zeta \in \mathbb R$ ($\mathfrak M_{\omega,r_0}=\{(\frac{3r_S}{2},\varphi),\varphi \in [0,2\pi]\}$, all circular orbits are outside the complex magnetic monopole which is identified with the photon sphere). The adiabatic transport generates a phase difference between $\phi_+$ and $\phi_-$ which induces some interferences in the quantum teleportation explaining the small oscillations in the fidelity. 

\subsubsection{Geodesics reaching the event horizon:}
\begin{figure}
\begin{center}
\includegraphics[width=7cm]{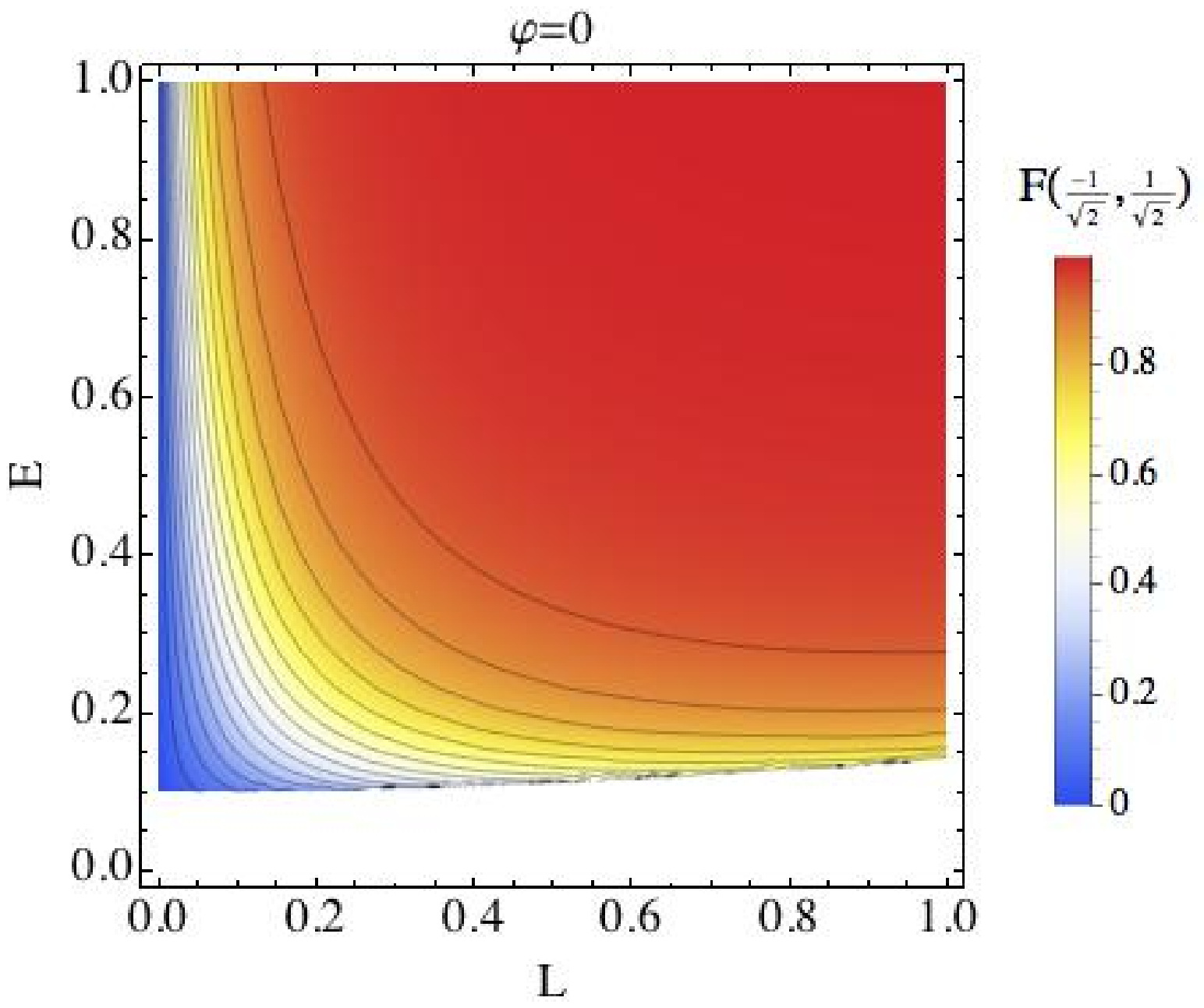}\includegraphics[width=7cm]{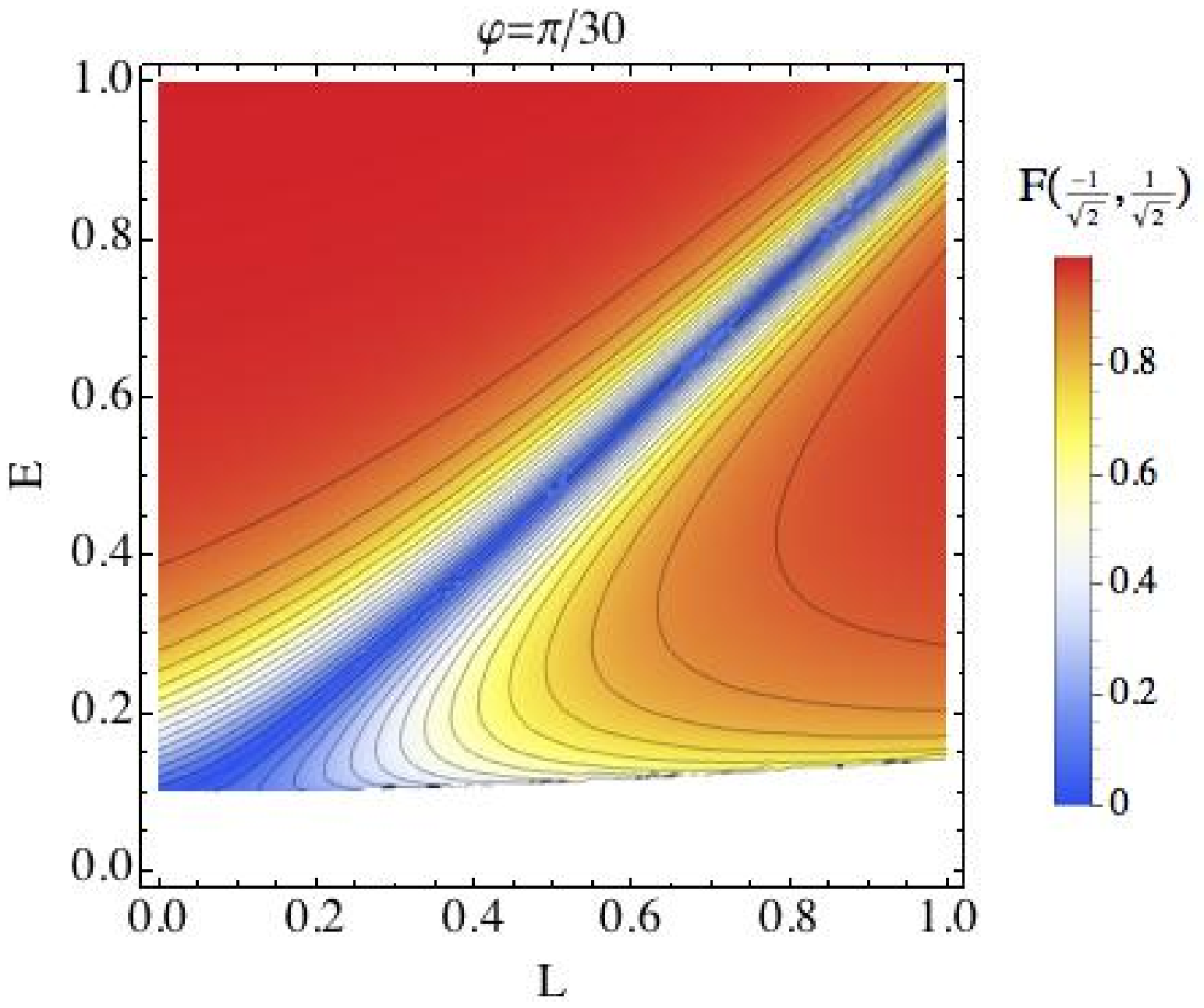}\\
\includegraphics[width=7cm]{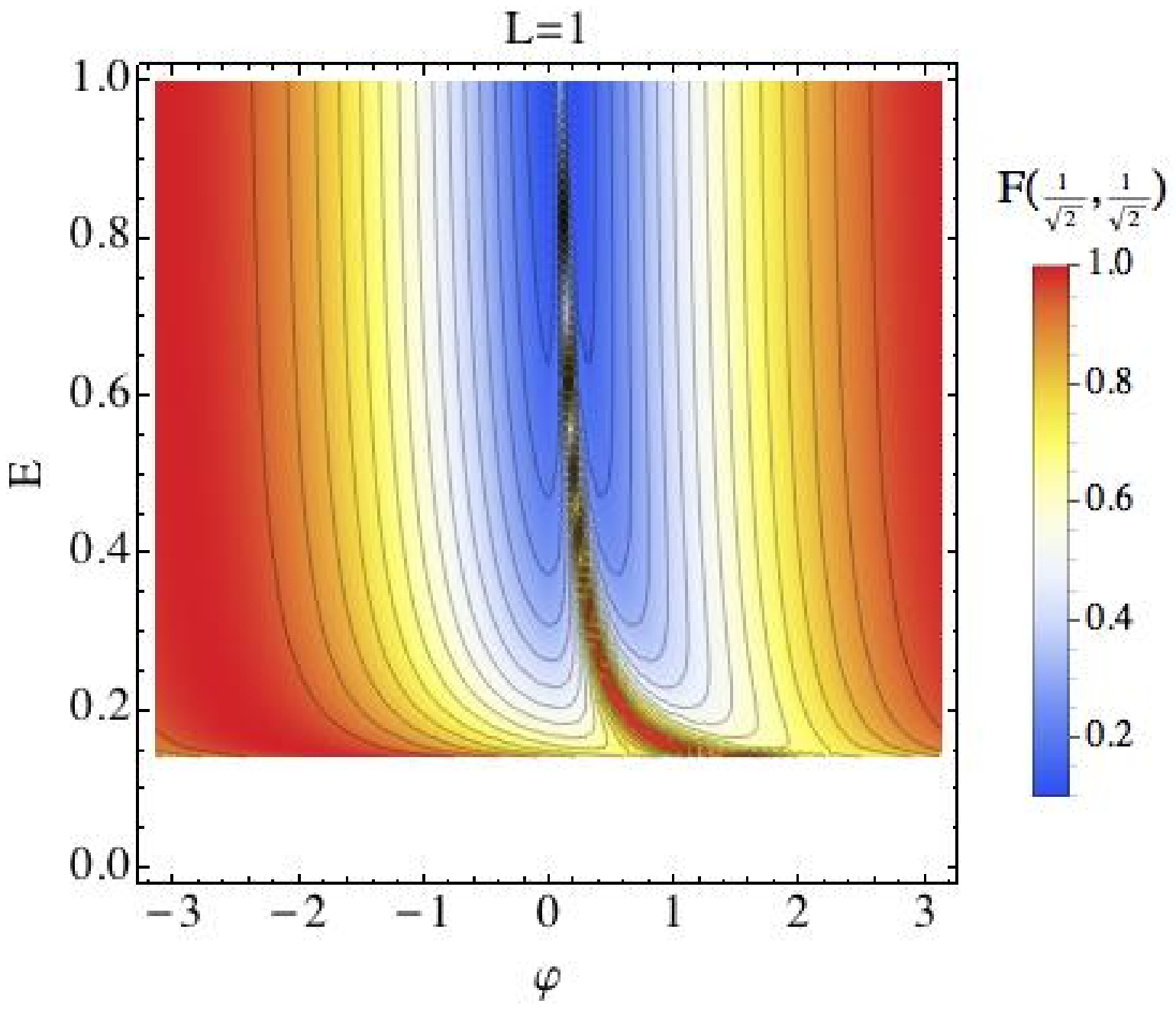}\includegraphics[width=7cm]{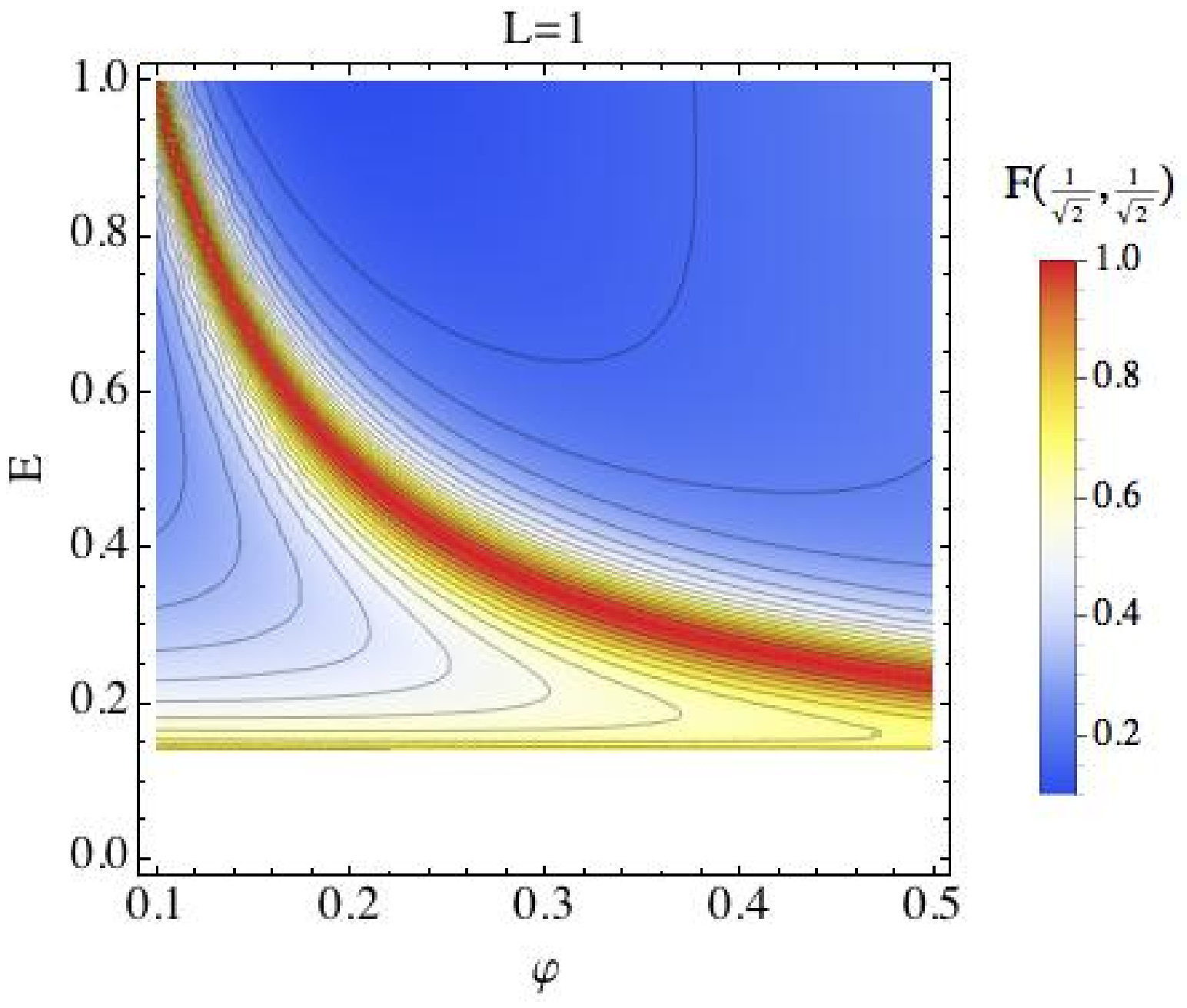}\\
\includegraphics[width=7cm]{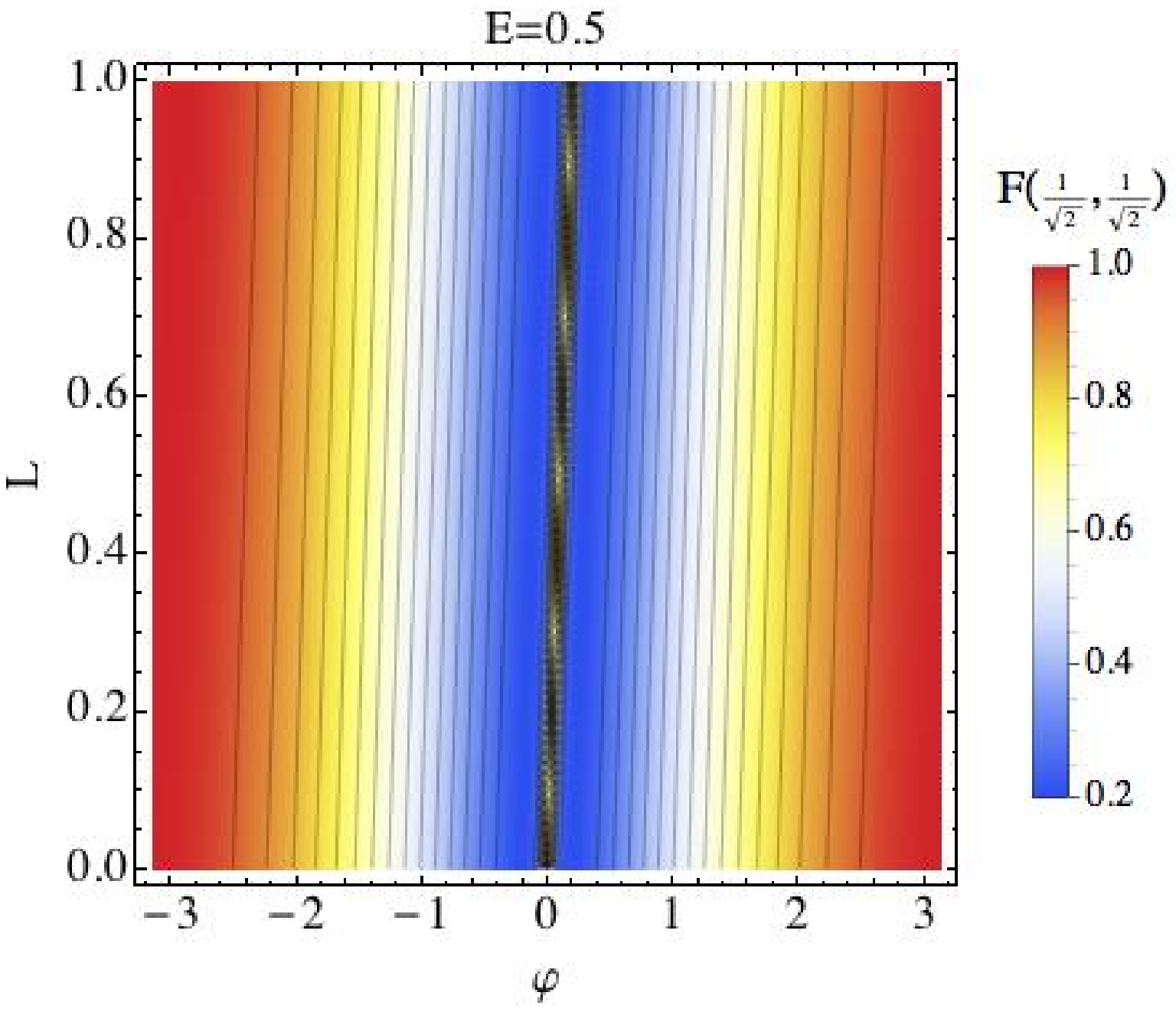}\includegraphics[width=7cm]{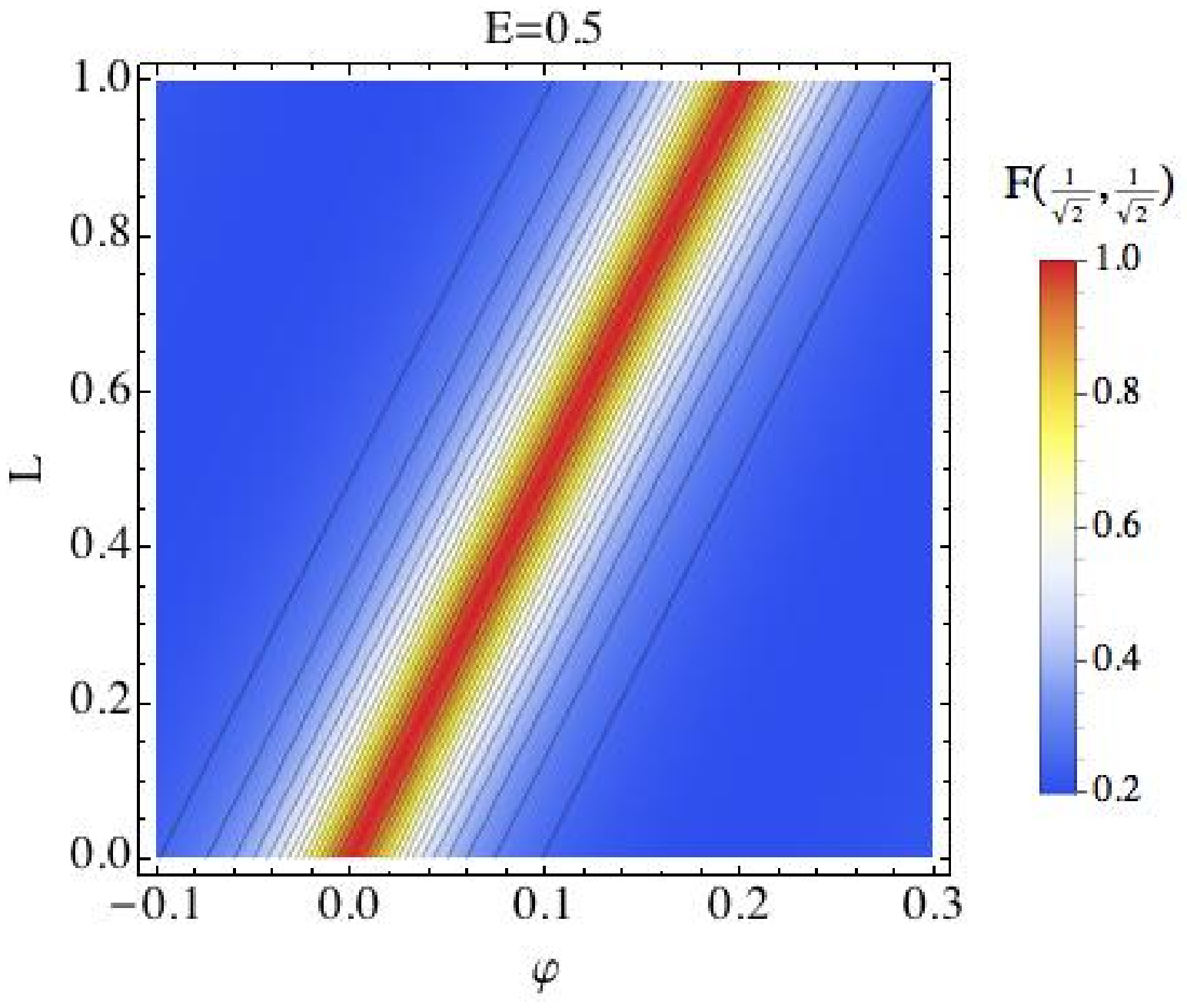}\\
\caption{\label{fidelity_schwarz_HE}Fidelity $F(\alpha,\beta)$ (eq. \ref{fidelity}) of the EPR teleportation protocol, for Alice following a geodesic almost reaching the event horizon and Bob comoving with the black hole, with respect to $E$ and $L$ (first integrals of the geodesic) and $\varphi$ the angular position of Alice when she almost reaches the event horizon (Alice being started from $\varphi_0=0$), its final radial position $r_f$ being such that $\frac{r_f-r_S}{r_S} = 10^{-2}$. Alice realizes its part of the protocol when she reaches its final point near the event horizon.}
\end{center}
\end{figure}

We consider geodesics starting far from the event horizon and almost reaching it by an adiabatic process for the qubit evolution. Since $\zeta \yrightarrow{r \to r_S} + \imath \infty$ we can suppose that $e^{\frac{\imath}{2} \int_0^{\tau_H-\epsilon} \zeta d\tau} \simeq 0$ ($\tau_H$ being the proper time needed to reach the event horizon and $\epsilon \ll 1$). Moreover $\phi_+ \yrightarrow{r \to r_S} \frac{1}{\sqrt 2} \left(1 \atop 1\right)$, $\phi_- \yrightarrow{r \to r_S} \frac{1}{\sqrt 2} \left(-1 \atop 1\right)$, and since $(u^A)_{A\in\{t,r,\theta,\varphi\}} = \left(\frac{E}{T},-\sqrt{\frac{E^2}{T^2}-1-\frac{L^2}{r^2}},0,\frac{L}{r}\right)$, we can compute an evaluation of the fidelity of the teleportation protocol for geodesics almost reaching the event horizon, see figure \ref{fidelity_schwarz_HE}.
The fidelity oscillates with the relative angular position of Alice when she reaches the event horizon (for the teleportation of a ``Schr\"odinger cat'', no oscillation occurs for the teleportation of $|0\rangle$ or $|1\rangle$). The dependence from $E$ and $L$ is small except for the small values of these first integrals.

%\subsection{Linearized Kerr black hole}
%We consider the metric associated with rotating black hole with slow rotation which is characterized by the metric:
%\begin{equation}
%d\tau^2 = T(r)^2dt^2 - R(r)^{-2}dr^2 - r^2d\Omega^2 + K(r)T(r) \sin^2 \theta dt d\varphi
%\end{equation}
%($d\Omega^2 = d\theta^2+\sin^2\theta d\varphi^2$). The linearized Kerr metric is obtained with $R(r)=T(r) = \sqrt{1-\frac{r_S}{r}}$ and $K(r)=\frac{2r_S a}{rT(r)}$ ($a=\frac{J}{M}$, $J$ being the angular momentum of the black hole). The tetrad fields are $e^0 = Tdt+K\sin^2\theta d\varphi$, $e^1=R^{-1}dr$, $e^2=rd\theta$ and $e^3=-\sqrt{K^2\sin^4\theta+r^2\sin^2\theta} d\varphi$; and the non-zero components of the Lorentz connection are $\omega^{01}=-RT'dt+\frac{R(KT'-K'T)\sin^2\theta}{T+1}d\varphi$, $\omega^{02}=-\frac{K}{2r} \sin(2\theta) d\varphi$, $\omega^{03}=-\frac{K'+KT'}{(T+1)\sqrt X} \sin^2 \theta dr - \frac{K}{2\sqrt X} \sin(2\theta) d\theta$, $\omega^{12} = Rd\theta$, $\omega^{13}=\frac{RT(K'+KT')}{(T+1)\sqrt X} \sin^2 \theta dt - \frac{RK(TK'-T'K)\sin^4\theta+R(T+1)r\sin^2\theta}{(T+1)\sqrt X} d\varphi$, and $\omega^{23}=\frac{KT}{2r\sqrt X} \sin(2\theta) dt - \frac{2K^2\cos \theta \sin^3\theta+r^2 \sin(2\theta)}{2r\sqrt X} d\varphi$ (with $X=K^2\sin^4\theta+r^2\sin^2\theta$).

\section{Conclusion}
A localized qubit in general relativity is described (at the adiabatic limit) by a geometric structure including the description of the quantum states and of the spacetime geometry, i.e. the composite bundle $\mathbf{\Phi}_+ \oplus \mathbf{\Phi}_- \to \mathcal A \times TM \to M$ where $\mathcal A$ is the space of Lorentz connections. In this bundle, inside a particular submanifold (the complex magnetic monopole $\mathfrak M_{\mathcal A}$) decoherence processes appear on the qubit. This complex magnetic monopole is a sphere surrounding the event horizon for a Schwarzschild black hole, with radius decreasing with the increase of the angular momentum of the qubit (it is infinite for $L \leq \frac{r_SE}{2}$, is equal to $\frac{3r_S}{2}$ (the photon sphere) for the circular orbits, and tends to $r_S$ (the event horizon) with $L \to + \infty$). We have two different decoherence processes, a dynamical decoherence associated with the non-unitary dynamical phases and a geometric decoherence associated with the non-unitary geometric phases (and depending only on the shape of the followed path and not from the proper time). The physical origin of these processes is related in a Rindler spacetime to the Unruh radiation and we can then postulate that in the general case it is related to the Hawking radiation (since the Unruh effect can be considered as the near-horizon form of the Hawking radiation). We have shown how these decoherence processes degrade the fidelity of the quantum teleportation protocol if Alice falls to the event horizon, the adiabatic framework permitting to obtain a simple formula to compute this fidelity with respect to the spacetime position and to the four-velocity of Alice when she realizes her part of the protocol.\\
The approach of the adiabatic dynamics of localized qubits presented in this paper is valid in the context of the semi-classical approximations (WKB, no second quantization, adiabatic limit). But it permits to consider all spacetime geometry and all geodesics. In contrast, the approach of the Fuentes-Schuller Mann model does not make approximation in the quantum field theory but is restricted to the Rindler spacetime (neighbourhood of the event horizon of a Schwarzchild black hole). The two models are then complementary. The calculation of the quantum field of a qubit in a generic curved spacetime with a strong localization without any approximation is a very difficult problem. The two approaches permitt to have complementary views of the problem with simple calculations.

\end{document}